\documentclass[aps]{revtex4}

\pdfoutput=1
\usepackage{graphicx}

\begin{document}

\title{On the Concept of Dynamical Reduction : The Case of Coupled Oscillators}

\author{Yoshiki Kuramoto$^{1}$ and Hiroya Nakao$^{2}$}

\address{$^{1}$Department of Physics, Kyoto University, Kyoto 606-8502, Japan (emeritus)\\
$^{2}$Department of Systems and Control Engineering, Tokyo Institute of Technology, Tokyo 152-8552, Japan\\
}


\begin{abstract}
An overview is given on two representative methods of dynamical reduction known as {\it center-manifold reduction} and 
{\it phase reduction}. These theories are presented in a somewhat more
unified fashion than the theories in the past. 
The target systems 
of reduction are coupled limit-cycle oscillators. 
Particular emphasis is placed on the remarkable structural similarity existing between these theories. 
While the two basic principles, i.e. (i) reduction of dynamical degrees of freedom and (ii) transformation of reduced evolution equation to a canonical form, are shared commonly by reduction methods in general, it is shown how these principles are incorporated into the above two reduction theories in a coherent manner. 
Regarding the phase reduction,
a new formulation of perturbative expansion is presented for discrete populations of oscillators. 
The style of description is intended to be so informal
that one may digest, without being bothered with 
technicalities, what has been done after all
under the word \textit{reduction}. 
\end{abstract}

\maketitle

\section{Introduction}

The dynamics of nonlinear dissipative systems are most commonly
modeled mathematically with coupled ordinary and partial differential equations~\cite{Strogatz}.
The specific forms of evolution equations are often too complex to handle or may even be too poorly known. 
Thus, the effort of reducing them to a much simpler form by cutting off
unessential parts would be extremely important to the understanding of 
nonlinear dynamical systems. 
We will call such work of purification
 \textit{dynamical reduction} or simply
\textit{reduction}. \par

\maketitle

It seems widely accepted today that the concept of reduction has worked as a 
strong motivating force for the advancement of nonlinear science over the past half century. 
As a remarkable fact, reduction provides, besides the
explanation of how this or that dynamical behavior observed occurs, 
a powerful tool for predicting utterly new types of behavior 
whose existence in nature has never been imagined before. 
Since the result of reduction
often takes a universal form, the predicted dynamics is 
also expected to appear universally. 
The universal form of the reduced
evolution equation also implies that seemingly unrelated phenomena
occurring in systems of completely different physical constitution
may have something in common at a deeper mathematical level. 
Accumulation of such findings, each of which may be a small one,
may ultimately lead to our refreshed view of nature. 
\par

Reduction is generally practicable when the dynamics of largely separated time scales 
coexist in the system of concern, that is, when dynamical variables of fast motion and those
of much slower motion are coupled. 
Traditionally, two basic ideas have been known
for the reduction of the dynamics of this type of systems. 
One is the idea of the so-called \textit{adiabatic elimination} and the other is that of \textit{averaging}. \par
In adiabatic elimination, rapidly relaxing variables are eliminated by 
considering that they
adiabatically follow the motion of slow variables.  
Consequently, the 
system's dynamics comes to be
confined practically to a lower dimensional subspace spanned only by 
a small number of slow variables. 
The broad
applicability of this theoretical tool was emphasized by Hermann Haken under the term 
coined by him 
\textit{slaving principle}, i.e.$\;$ the principle that slow modes slave fast modes~\cite{Haken1,Haken2}. \par
In contrast to adiabatic elimination, the method of averaging is irrelevant to the reduction of dynamical degrees of freedom. 
This method can be applied typically when rapid oscillatory motion appears parametrically in the evolution equation governing the slow variables. 
Since such fast parametric modulation should have minor effects on the
slow dynamics, it may safely be removed by time-averaging. 
Averaging represents a crucial
tool in the asymptotic theory of weakly nonlinear oscillations developed extensively
by Russian scientists such as Krylov, Bogoliubov and Mitropolsky~\cite{Bogoliubov}. 
We shall see in later
sections that averaging is equivalent to slightly changing the definition of slow variables. \par
In this article, confining our concern to limit-cycle oscillator systems~\cite{GuckenheimerHolmes,Winfree0,Kuramoto,Ermentrout0,Hoppensteadt,Pikovsky},
we show how the above-described two traditional ideas of reduction can be incorporated more systematically into the form of a single
asymptotic theory. 
Note that,
in either adiabatic elimination or averaging, reduction is achieved most satisfactorily when the system involves special degrees of freedom whose time evolution is extremely slow or, to put it differently, their stability is nearly neutral. 
In the systematic reduction theory of our concern, which is a kind of a perturbation theory, the unperturbed part is chosen so as to include 
modes with perfect neutrality. 
The actual system is supposed to be
a perturbed system in which 
the weak perturbation would cause a slow evolution of the otherwise neutral modes. \par
In order that such a reduction theory may enjoy a universal applicability,
these neutral modes should also be of a universal nature.
In our macroscopic and mesoscopic world, we know at least two universal situations
where neutrally stable modes emerge. The first is met at the critical point of 
bifurcation where the critical eigenmodes are neutral in stability.
The second is the situation where a certain continuous symmetry 
inherent in the system has been broken, producing a neutrally stable mode
{\footnote{As the third possibility, conserved quantities play the
role of neutral modes, because these quantities, once changed, never
recover their original value. 
The Enskog-Chapman theory~\cite{Chapman} of deriving
hydrodynamic equations from the Boltzmann gas kinetic equation gives
an important example of reduction theory that makes full use of the
neutrality of conserved quantities. }.
The second type of neutral mode may be called \textit{phase} in a broad sense.
The representative reduction theories making full use of these neutral 
modes are known, respectively, as \textit{center-manifold reduction} and 
\textit{phase reduction}~\cite{Kuramoto,Winfree0,GuckenheimerHolmes,Hoppensteadt,Ermentrout0}.
\par
Fortunately, each of these theories is applicable to 
limit-cycle oscillator systems because in many cases the oscillation is a result of the Hopf bifurcation and the oscillation itself breaks the translational symmetry of time.
The science of coupled oscillators is becoming a rapidly expanding
field today. 
Since the reduction of coupled oscillator models provides a powerful tool for many practical purposes, reexamining its theoretical basis would be of considerable value. 
\par
Contrary to our intuition, 
the reduction theory for a single oscillator system and that for multi-oscillator
systems can be formulated quite similarly~\cite{Kuramoto}. 
In the center-manifold reduction, the 
linearized system about the fixed point right at the Hopf bifurcation
is chosen as the unperturbed system. 
The perturbation includes 
nonlinearity and the deviation from the bifurcation point, both of which
are assumed to be sufficiently small. 
When we proceed to systems of
weakly coupled oscillators, the coupling force from the other oscillators will simply be regarded as a sort of perturbation acting on the oscillator in question. \par
In the phase reduction, the unperturbed system is given by a single
limit-cycle oscillator running along or near the closed orbit. 
There the phase of the oscillatory motion represents a neutrally stable degree of freedom. 
Similarly to the oscillator near the Hopf bifurcation, weak coupling force
from the other oscillators and weak external drive, if necessary, 
are treated as perturbation. \par
The fact that the unperturbed system contains a neutral mode implies that
its solution as $t\rightarrow\infty$ includes at least one arbitrary parameter.
For instance, the solution of the linearized system at the Hopf bifurcation 
as $t\rightarrow\infty$ is given by a harmonic oscillation $\dot{z} =i\omega z$,
where the complex amplitude $z$ may be multiplied by an {\it arbitrary} complex
number. 
For a limit-cycle motion along the closed orbit, the phase $\phi$
changes as $\dot\phi=\omega$ or $\phi=\omega t+\psi$, where the
initial phase $\psi$ is {\it arbitrary}. \par
The most basic idea of reduction is to
reinterpret such an arbitrary constant as a dynamical variable, and absorb
the effect of perturbation into this variable thereby causing its slow 
evolution.
At the same time, the unperturbed form of the solution must inevitably
be changed, but only a little. This small change modifies the dynamics of the slow variable, which in turn produces an even
smaller correction of the solution form. 
Such processes 
of mutual correction between the solution form and the dynamics of the
slow variables will repeat indefinitely. 
Still, this is not our goal of
reduction.\par
In the evolution equation of the slow variable obtained at each stage of approximation, fast parametric 
oscillations will generally appear. 
Such a parametric modulation, 
which is
much faster than the slow dynamics of our concern, could be removed by time-averaging. 
The same simplification of the evolution equation could be achieved more systematically through a variable transformation called near-identity transformation.
\par
Our goal is to put the whole of the above stated ideas into a mathematical form. 
We shall see how this can be achieved in each of the two 
representative theories of reduction. \par
Section 2 and 3 will be devoted, respectively, to center-manifold reduction
and phase reduction. 
A few comments supplementary to these sections together with a general remark on the significance of reduction in nonlinear science will be given in the final section.
In the Appendix, some recent developments in the phase-reduction theory, such as the phase-amplitude reduction, Koopman-operator approach, and extension to non-conventional dynamical systems, are briefly described.
Throughout the present paper, we only present the main line of the theory, ignoring all technical details. 

\begin{figure}[!h]
\centering
\includegraphics[width=0.75\hsize,clip]{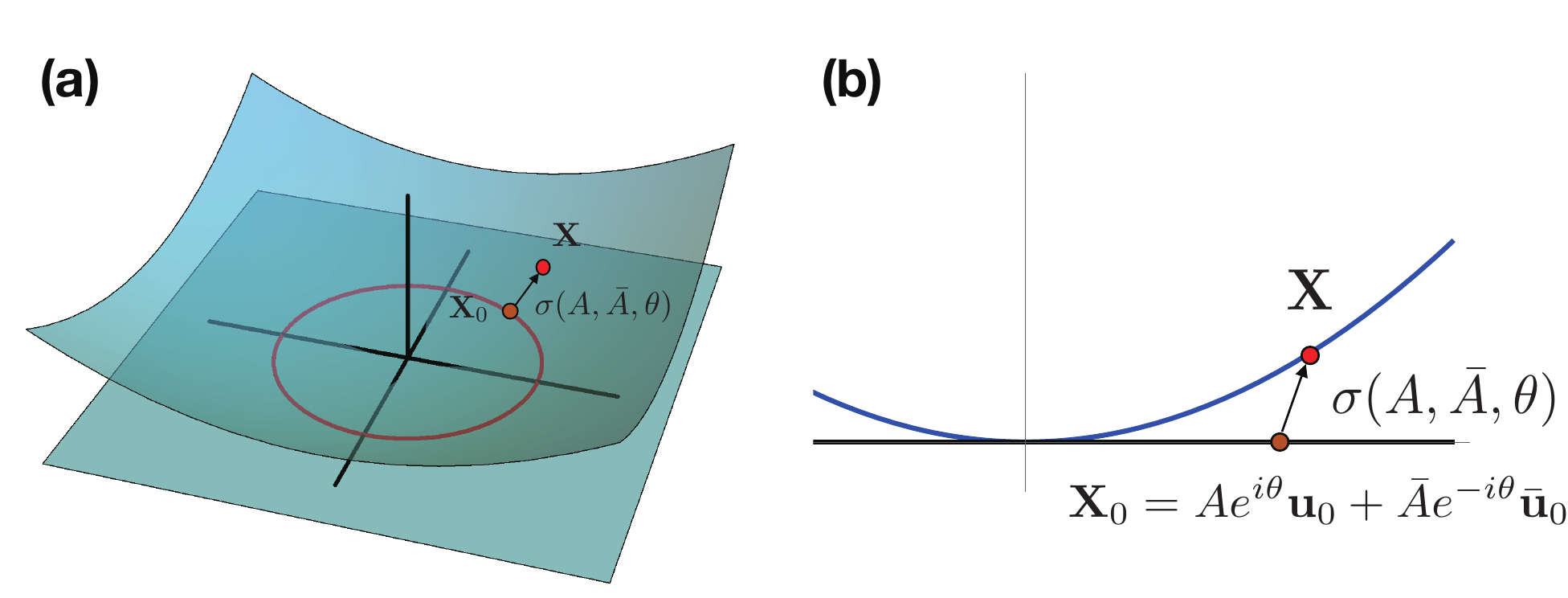}
\caption{
Schematic figures of a neutral solution and a center manifold. (a) The neutral solution ${\bf X}_0$ in the $2$-dimensional plane spanned by ${\bf u}_0$ and $\bar{\bf u}_0$, characterized by the complex amplitude $A$ and phase $\theta = \omega t$. The center manifold is tangent to this plane at ${\bf X} = 0$. (b) Representation of the true solution ${\bf X}$ on the center manifold by using the neutral solution ${\bf X}_0$ and a small correction $\sigma(A, \bar{A}, \theta)$.
}
\label{fig1}
\end{figure}

\section{Center-manifold reduction}

We begin with the center-manifold reduction for a single oscillator, and then proceed to multi-oscillator systems.
It will be seen how the two elements of reduction, namely, the reduction of dynamical degrees of freedom and transformation of the evolution equation to a canonical form,
are incorporated into a systematic perturbation theory. 

\subsection{Reduction of a single free oscillator}

The situation of our concern is the neighborhood of the Hopf bifurcation of an $n$-dimensional dynamical 
system $\dot{\bf X}={\bf F(X)}$ with fixed point ${\bf X}=0$. 
We separate
$\bf F$ into an unperturbed part and perturbation, where the former 
is chosen to be the linear part right at the bifurcation, and the latter 
includes the remainder, that is, the system's nonlinearity and the effect of small deviation
from the bifurcation point. 
Thus, the system to be reduced is written as
$$
\dot{\bf X}={\hat L}_0{\bf X}+\epsilon{\hat L}_1{\bf X}+{\bf N(X)}.
\eqno(2.1)
$$
For simplicity, the small deviation from the bifurcation point, represented by the second term on the right-hand side, is given 
only by the linear term with small bifurcation parameter $\epsilon$; the nonlinear
part of $\bf F(X)$ is abbreviated as $\bf N(X)$. 
These two terms 
constitute the perturbation acting on the linear system at the criticality.
The Jacobian ${\hat L}_0$ has a pair of pure imaginary eigenvalues $\pm i\omega$, and 
the corresponding eigenvectors are denoted as
${\bf u}_0$ and $\bar{\bf u}_0 $, respectively, where the bar denotes complex conjugate; 
the real part of all the other eigenvalues are assumed negative. 
Thus, the unperturbed system $\dot{\bf X}={\hat L}_0\bf X$ is neutrally stable. Its solution as $t\rightarrow\infty$, which is called the neutral solution,
is given by
$$
{\bf X}_0=Ae^{i\theta}{\bf u}_0+{\bar A}e^{-i\theta}\bar{\bf u}_0,
\eqno{(2.2)}
$$
where $\theta=\omega t$ 
and $A$ is an arbitrary complex amplitude. 
Note that $e^{i\theta}{\bf u}_0$
and its complex conjugate are the eigenfunctions of the operator 
${\hat L}_0-\omega\frac{d}{d\theta}$ with zero eigenvalue. This fact will be used later.\par
As was stated in the introductory section, the basic idea of reduction is to reinterpret the
arbitrary parameter $A$ appearing in the time-asymptotic solution 
as a dynamical variable and try to absorb the effect of perturbation
into the slow evolution of $A$. 
At the same time, the change of the
the solution form from (2.2), however small it may be, is inevitable. 
Thus, a small correction term must be added to the right-hand side of (2.2) to express the true solution.
The crucial assumption here is that this correction term is given by
a functional of the neutral solution. 
As a result, the true solution takes the form
$$
{\bf X}=Ae^{i\theta}{\bf u}_0+{\bar A}e^{-i\theta}{\bar{\bf u}}_0+{ \sigma}(A, {\bar A}, \theta). \eqno{(2.3)}
$$
By assumption, the dependence of the small correction term $\sigma$ on $\theta$ always 
appears as the combination $Ae^{i\theta}$ or ${\bar A}e^{-i\theta}$.
Since this time-asymptotic solution is parameterized with the two independent variables, i.e.$\,$the real and imaginary parts of $A$,
the dynamics takes place on a 2D subspace. 
This subspace is the result of a slight deformation of the 2D critical eigenspace, and is called mathematically
the center manifold~\cite{GuckenheimerHolmes}.
Figure~\ref{fig1} shows the neutral solution, center manifold, and representation of the true solution schematically.
\par

A few more remarks should be made on the above form of the true solution. 
Suppose first that the dynamical model to be reduced is 
already two-dimensional as exemplified by the FitzHugh-Nagumo oscillator 
and the Brusselator~\cite{Winfree0,Kuramoto}.
In such a case, the critical eigenspace itself spans the full phase space.
It may then seem that the correction term $\sigma$ is unnecessary,
and that all the effects of perturbation can completely be absorbed into
the slow evolution of $A$ and $\bar A$. 
This is true, but the 
existing theories of center-manifold reduction still assume the solution form equivalent to (2.3) without 
inquiring into the dimension of the dynamical system model.
There is actually no contradiction here, because the role played by
the term $\sigma$ is twofold. 
It certainly represents a slight deformation of the invariant space when the dimension of the dynamical 
system is higher than two. 
But what should never be overlooked is that 
$\sigma$ also changes the definition of $A$ slightly like 
$A\rightarrow A+s(A,{\bar A},\theta )$; when we have written down (2.3)
with nonvanishing $\sigma$, the meaning of $A$ has been changed
a little.
This transformation of $A$ gives an example of the 
near-identity transformation. 
In short, for higher dimensional
systems, $\sigma$ plays double roles, while for two-dimensional systems
it plays a single role.\par
While we say that variable transformation is necessary in order that the resulting evolution equation
for $A$ may take a canonical form, what is actually meant by \textit{canonical form}?
In the present reduction theory, it means that fast parametric oscillation
does not appear in the evolution equation for $A$, that is, 
$\theta$-dependence is excluded from that equation. 
Since $\theta$
always appears as the combination $Ae^{i\theta}$ or its complex conjugate,
$\dot A$ is required to assume the form
$$
{\dot A}=G(A,{\bar A})=H( |A|^2 )A \eqno{(2.4)}
$$
as long as G can be expanded in powers of $A$ and $\bar A$. 
The unknown quantities $G$ and $\sigma$ are 
small because they are nonvanishing only because the perturbation is nonvanishing . \par
Another point to be noticed about the solution form (2.3) is that
the separation between the neutral part and the correction term 
$\sigma$
is not unique without imposing some additional condition. 
To make this separation unique, we should remember that
$e^{i\theta}{\bf u}_0$ and $e^{-i\theta}\bar{\bf u}_0$ are the zero
eigenfunctions of ${\hat L}_0 - \omega \frac{d}{d\theta}$. 
Thus, as the most natural choice,
we exclude these components from $\sigma$; still $\sigma$
may include the vector components ${\bf u}_0$ and $\bar{\bf u}_0$,
but only in the combination with higher harmonics 
$e^{im\theta} (m\ne \pm 1)$. 
This method of separation is crucial to the formulation of the present reduction theory, because, as we see below,
 it enables us to achieve the two tasks, i.e.$\:$ reduction of the degrees of freedom and 
transformation of the reduced equation into a canonical form, at one
stroke rather than stepwise.\par
We are now left with the two unknown quantities
$G$ and $\sigma$, and try to find them perturbatively by inserting
(2.3) into (2.1) and using (2.4). 
Our direct concern is the functional form of $G$ which determines
the reduced evolution equation; $\sigma$ is necessary only for finding 
small, but not always negligible, corrections to $G$ beyond the lowest-order approximation.\par
When we refer to a perturbation theory, some small parameter usually appears
explicitly in the theory. 
Although a small parameter $\epsilon$ appears in (2.1) measuring the small distance from the bifurcation point, this smallness causes 
small-amplitude oscillations, so that the amplitude $\bf X$ or $A$
should also be treated as a small quantity.
This is, so to speak, an
{\it implicit} small parameter. 
How the smallness of $A$ is related to
$\epsilon$ is not known in advance. 
In some existing reduction theories,
however, the smallness $\epsilon^{1/2}$ is assigned to $A$ {\it a priori}, but such an assignment 
should be justified only {\it a posteriori}. 
Therefore, we have to
proceed for the moment only with the assumption that the smallness of
$\epsilon$ and that of $A$ are independent. 
Similarly,
the smallness of $G$ and $\sigma$ should also be regarded as
implicit small quantities whose smallness is independent of $\epsilon$
at this stage of the theory.\par
We now substitute (2.3) into (2.1). 
Since the time derivative is only through $A$, $\bar A$ and $\theta$, $d/dt$ is replaced with
$\omega\frac{d}{d\theta} +G\frac{d}{dA}+{\bar G}\frac{d}{d{\bar A}}$. 
The equation obtained in this way can be expressed in the concise form
$$
Ge^{i\theta}{\bf u}_0+{\bar G}e^{-i\theta}{\bar{\bf u}}_0
-\biggl({\hat L}_0-\omega\frac{d}{d\theta} \biggr){\sigma}={\bf B}(A, {\bar A},\theta). \eqno{(2.5)}
$$
In the above equation, $\bf B$ is dominantly composed of the perturbation terms 
$\epsilon{\hat L}_1{\bf X}+{\bf N(X)}$ with $\bf X$ replaced with (2.3), plus,
less importantly, the time-derivatives of $\sigma$ through $A$ and 
$\bar A$, e.g.$\:$the term 
 $-G\frac{\partial\sigma}{\partial A}$ and its complex conjugate, are also included. \par
We now pretend that $\bf B$ is a known quantity.
Then, (2.5) formally represents a linear equation for the
unknowns $G,\bar G$ and ${\sigma}$ with the inhomogeneous term 
$\bf B$. 
Noting that $e^{i\theta}{\bf u}_0$ and its complex conjugate appearing
on the left-hand side are the zero eigenfunctions of ${\hat L}_0-\omega\frac{d}{d\theta}$, while by assumption $\bf \sigma$ is free from these components, we can find the unique solution to this linear equation.
This is achieved 
by decomposing $\bf\sigma$ into various eigenfunctions $e^{im\theta}({\bf u}_l, {\bar{\bf u}_l})$ of ${\hat L}_0-\omega\frac{d}{d\theta}$ with no-zero eigenvalues, where $m$ is an integer and ${\bf u}_l$ denotes the $l$-th eigenvector of $\hat{L}_0$. 
The equation for the zero eigenfunction components determines $G$ 
and $\bar G$. 
The result is given by
$$
 G=\frac{1}{2\pi}\int_0^{2\pi}e^{- i\theta}{{\bf u}_0^{\ast}} {\bf B}(A,{\bar A},\theta)d\theta,
\eqno{(2.6)}
$$
where ${{\bf u}_0^{\ast}}$ is the left eigenvector of ${\hat L}_0$ with eigenvalue $i \omega$, and hence,
$e^{- i\theta}{{\bf u}_0^{\ast}}$ is the zero eigenfunction of the 
adjoint operator of 
${\hat L}_0 - \omega \frac{d}{d\theta}$. 
We have assumed that the left and right eigenvectors of $\hat{L}_0$ are biorthonormalized as ${\bf u}_l^* {\bf u}_m = \delta_{l, m}$.
\par
Up to this point, we have pretended that $\bf B$ is a known quantity.
Actually, however, $\bf B$ includes the unknowns $G$ and $\bf\sigma$. 
Still the problem can be solved iteratively. 
The starting point is to 
neglect $G$ and $\sigma$ appearing in $\bf B$ as
relatively small. 
Then, (2.5) becomes
a truly linear inhomogeneous equation from which $G$ and $\bf\sigma$ are determined. 
These solutions
 are then inserted into 
$\bf B$. By using this improved $\bf B$, improved $G$ and $\sigma$
are obtained. 
Such processes can be repeated indefinitely, by which
$G$ and hence the reduced evolution equation is derived.
It takes the form of a power series expansion in $A$ like 
$$
{\dot A}=\alpha_0 A+\alpha_1 |A|^2 A+\alpha_2 |A|^4 A+\cdots,
\eqno{(2.7)}
$$
where $\alpha_0=O(\epsilon)$ and $\alpha_l=O(1)\: (l=1,2,\cdots)$
are complex coefficients.
\par
Although all calculational details are omitted here, it should be pointed out that there is no precise correspondence between the iteration step and the 
resulting power of $A$. 
For instance, when $\bf N(X)$ includes quadratic and cubic nonlinearity,
the contribution to the cubic term $|A|^2 A$ in the $\dot A$ equation comes not only 
from the lowest-order $\bf B$ but also from the corrected $\bf B$ 
resulting from the first iteration.
Thus, nonvanishing $\sigma$ is responsible for the correct expression of the cubic nonlinearity 
in the $\dot A$ equation. \par
If we drop in (2.7) all the nonlinear terms higher than the cubic term, the
so-called Stuart-Landau equation is obtained. 
We find that the solution to this equation takes the
scaling form $A(t)=\epsilon^{1/2}{\tilde A}(\epsilon^2 t)$, implying that 
the first three terms on the right-hand side of (2.7) have the same order of magnitude.
Admittedly, the condition $\Re \alpha_1<0$ must be satisfied so 
that $|A|$ 
may not escape to infinity. 
Under this condition or, in other words, if 
the Hopf bifurcation is of a supercritical type, then
the contribution from the terms higher than the cubic 
turns out negligibly small.
Thus, unlike usual perturbation theories, the meaningful
truncation of the perturbation expansion does not result
automatically,
but some arguments of physical flavor are required.\par

\subsection{Reduction of a perturbed oscillator and coupled oscillators}

The above arguments leading to the well known one-oscillator model 
would not be of much practical interest. 
Moreover, for such a one-oscillator 
reduction problem, far more precise mathematical arguments could be
developed~\cite{GuckenheimerHolmes}. 
In contrast, for large systems of coupled oscillators, which are our main target of study, no mathematical theory equally rigorous to the one-oscillator theory seems to exist. 
Therefore, it seems 
difficult to proceed further without relying upon some 
unverifiable intuition. \par
We consider first diffusively coupled homogeneous fields of
oscillators with sufficient spatial extension described typically with 
reaction-diffusion equation $\dot{\bf X}={\bf F(X)}+{\hat D}
\nabla^2{\bf X}$, where ${\bf X}$ is now a function of spatial coordinate ${\bf r}$ and time $t$, and ${\hat D}$ is a matrix of diffusion coefficients. 
Mathematically, how to do reduction for such systems may seem
desperately difficult as compared with the reduction of a single
oscillator. 
This difficulty could be imagined, e.g.$\:$from the fact 
that we do not
have a clear concept of infinite-dimensional center manifold.
Incidentally, the naming ``center-manifold reduction'' commonly used may therefore be inappropriate for many-oscillator problems. 
Instead, we sometimes used the term ``reductive perturbation'' in the past, although this term will not be used in this article.\par
From a physicist's point of view, reduction of reaction-diffusion systems
can be achieved only by slightly modifying the reduction of a single
oscillator dynamics. 
Let us confine our arguments to reaction-diffusion equation. 
The key idea is that we still treat the problem as a 
one-oscillator problem by regarding the diffusion coupling 
${\hat D}\nabla^2{\bf X}$ as a weak
perturbation acting on a local oscillator. 
Here again, the assumed weakness
of the diffusion coupling could only be justified {\it a posteriori} by making use of what is expected physically that the length scale of the  spatial variation of
$\bf X$ will become increasingly longer as the 
Hopf bifurcation is approached. \par
We try to find the solution of the problem by generalizing the form of
(2.3) and (2.4). 
The only difference from those equations is that 
$A$ and $\bar{A}$ are now functions of ${\bf r}$ and $t$ and
the unknown quantities $G$ and $\sigma$ are assumed to depend also on various spatial derivatives of $A$ and $\bar A$. 
These spatial derivatives may
be looked upon as infinitely many parameters parameterizing the 2D
center manifold of a local oscillator. 
How to solve the problem is almost the same as the above. 
An equation similar to (2.5) is obtained, where the $\bf B$ term
includes, most importantly, ${\hat D}\nabla^2{\bf X}$, and less important terms arising from the time derivatives of $\bf\sigma$ through 
$A$, $\bar A$ and their various spatial derivatives.
$G$ and $\sigma$
can be found iteratively again by starting with the $\bf B$
in the lowest-order approximation 
which is free from the unknowns $G$ and $\sigma$. 
\par
In the resulting $\dot A$ equation, the new terms 
$\nabla^2 A$ and various spatial derivatives of $A$ higher than the second order appear. 
To see how to find a meaningful truncation of this series expansion,
we have to resort to a scaling argument again. 
The assumed scaling
form of $A$ includes spatial scaling as 
$A(t, {\bf r})=\epsilon^{1/2}{\tilde A}(\epsilon^2 t, \epsilon^{1/2}{\bf r})$.
As a result, we find that the most dominant new term to be included in the
 $\dot A$ equation
 is the diffusion term
$\beta\nabla^2 A$. 
The diffusion coefficient $\beta$ is a complex number given by 
$\beta={\bf u}^{\ast}_0{\hat D}{\bf u}_0$. 
In this way, the nontrivial lowest-order
equation becomes the so-called complex Ginzburg-Landau equation~\cite{Kuramoto}
$$
{\dot A}=\alpha_0 A+\alpha_1 |A|^2 A+\beta\nabla^2 A,\eqno{(2.8)}
$$
which in turn proves its consistency with the above-assumed scaling form. 
\par
Once the Stuart-Landau equation has been established as the nontrivial lowest-order evolution equation for $A$, the complex Ginzburg-Landau
equation results from a much easier argument than the above: 
First,
replace the diffusion coupling ${\hat D}\nabla^2{\bf X}$ with its approximation
 ${\hat D}\nabla^2{\bf X}_0$. 
Then, take its inner product with ${\bf u}^{\ast}_0 e^{-i\theta}$, i.e.$\:$ the zero
eigenfunction of the adjoint of ${\hat L}_0-\omega\frac{d}{d\theta}$. Add the
resulting term to the right-hand side of the Stuart-Landau equation,
producing the complex Ginzburg-Landau equation. 
Any other types of
perturbation than the diffusion coupling can be treated in a similar manner. As long as the lowest-order effect of the perturbation is concerned, each perturbation can be treated independently. 
In this way, 
many variants of the Stuart-Landau or complex Ginzburg-Landau equation could be obtained. 
The practical significance of such equations is great. 
For instance, besides the diffusion coupling,
weak periodic forcing with nearly resonant frequency may be introduced;
weak spatial heterogeneity may also be taken into account; diffusion coupling could be replaced with weak nonlocal coupling, etc~\cite{Kuramoto,Gambaudo,Mikhailov,Tanaka,Pikovsky,Nakao3,Bat}.

\section{Phase reduction}

We now proceed to another representative theory of reduction
called phase reduction, whose broad applicability to large populations and
networks of coupled oscillators is well recognized today~\cite{Winfree0,Kuramoto,Ermentrout0,Hoppensteadt}.
\par
In contrast to the center-manifold reduction, phase reduction is applicable to general large-amplitude oscillations far from the bifurcation point. 
Consider a well-developed limit-cycle oscillator described by a general $n$-dimensional dynamical system $\dot{\bf X}={\bf F(X)}$. Its periodic motion along the closed orbit itself is regarded as an unperturbed motion
to start with. Admittedly, no analytical solution of such a general oscillation is available. Historically, this fact might have worked as an obstacle to hitting on the idea that this intractable mathematical object can still be chosen as the starting point of a perturbation theory of reduction. This may partly explain the reason why the appearance of the phase reduction theory is relatively new compared e.g.$\:$with the asymptotic theory of weakly nonlinear oscillations.

\begin{figure}[!h]
\centering
\includegraphics[width=0.75\hsize,clip]{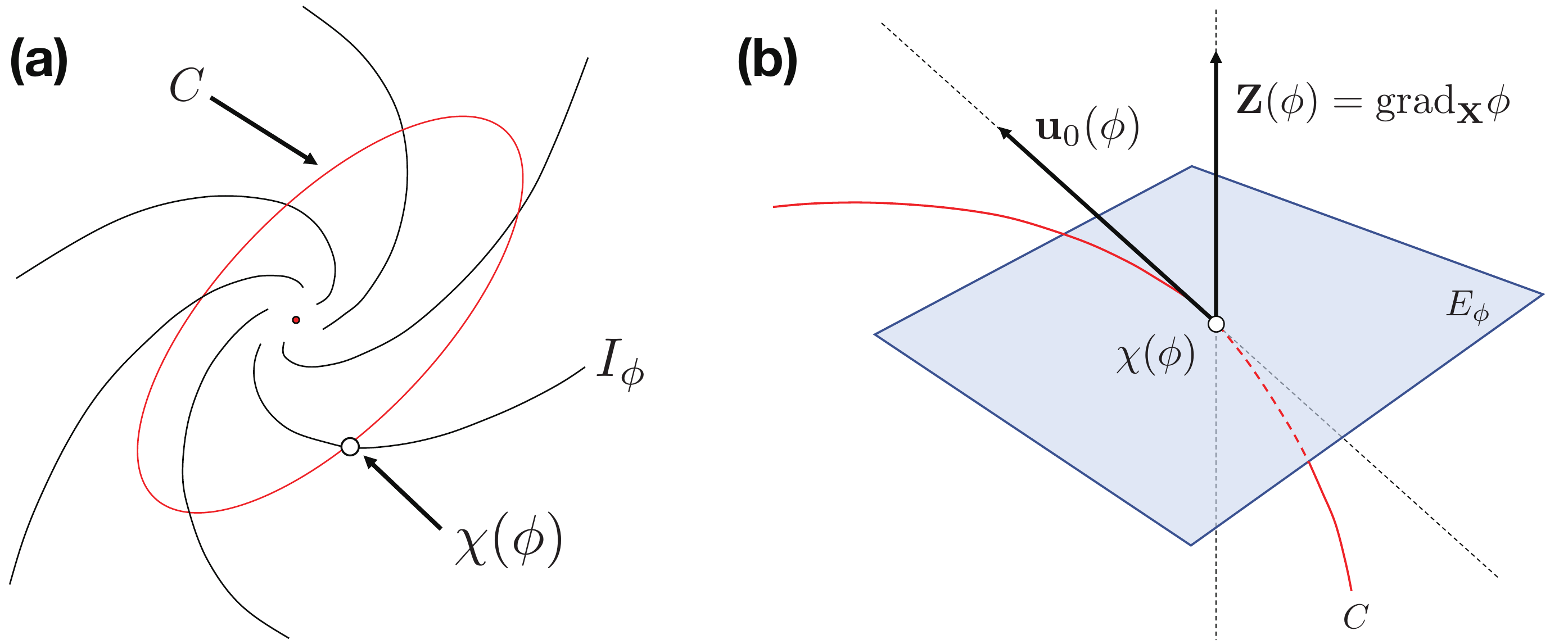}
\caption{Schematic figures of a limit cycle and isophase hypersurfaces. (a) An isochron $I_\phi$ is a $n-1$-dimensional hypersurface, which intersects with the limit cycle $C$ at the point $\chi(\phi)$. (b) An isophase $E_\phi$ used in Method II is identified with the tangent space of $I_\phi$ at point $\chi(\phi)$.}
\label{fig2}
\end{figure}

\subsection{Definition of phase}

Let the time-periodic solution of $\dot{\bf X}={\bf F(X)}$ with frequency
 $\omega$ be given by a $2\pi$-periodic function 
$\chi(\phi)$, where the phase $\phi$ varies as $\phi=\omega t+\psi$. The initial phase $\psi$ is
an arbitrary constant.
As noted in the introductory section, when the system's continuous symmetry has been broken, an arbitrary parameter must appear in the mathematical expression of the symmetry-broken state. This fact is true of the present case,
because the limit-cycle oscillation breaks the translational symmetry in time,
resulting in an arbitrary parameter $\psi$.
 \par
In conformity to the basic idea of phase reduction, we reinterpret the arbitrary constant $\psi$
as a dynamical variable, and absorb the effect of perturbation into 
 $\psi(t)$ to cause its slow evolution. \par
Up to this point, the phase $\phi$ represents a one-dimensional coordinate
defined along the closed orbit $C$ in such a way that the free motion of the oscillator 
on $C$ satisfies $\dot\phi=\omega$.
When the oscillator is perturbed, however weak the
perturbation may be, 
it will generally be kicked off the closed orbit $C$. In order that the phase may still have a definite meaning, we have to extend its definition outside $C$. Although we only need 
to define it in the vicinity of $C$, let us begin with its global definition by defining a scalar field $\phi(\bf X)$ for a general point
$\bf X$ in the phase space~\cite{Coddington}. \par
The most natural and frequently used definition of $\phi(\bf X)$ is such that
the unperturbed motion of the oscillator identically satisfies 
${\dot\phi}=\omega$.
The phase space will then be filled with an infinite family of 
the $(n-1)$-dimensional hypersurface of equal phase. An isophase hypersurface thus defined is called the isochron~\cite{Kuramoto,Guckenheimer,Winfree0,Winfree}. 
The isochron
of phase $\phi$ will be denoted by $I_\phi$. 
\par
Since the time-evolution of the oscillator's phase $\phi(\bf X)$ is only through the evolution of $\bf X$,
our unperturbed oscillator satisfies 
$$
{\dot\phi}={\rm grad}_{\bf X}\phi\cdot{\bf F(X)}=\omega.
\eqno{(3.1)}
$$ 
The second equality of the above equation gives the identity to be satisfied by the scalar 
field $\phi(\bf X)$. 
Note that the vector ${\rm grad}_{\bf X}\phi$ is vertical 
to $I_\phi$ along which the slope of $\phi$ is the steepest. Equation
(3.1) may therefore be understood as the natural requirement that the rate of change
of $\phi$ must be given by how many isochrons are crossed by the
oscillator per unit time. 
Figure~\ref{fig2}(a) schematically shows a limit cycle and isochrons. 
\par

\subsection{Method I }

Our goal is to reduce the perturbed equation 
$\dot{\bf X}={\bf F}({\bf X})+{\bf P}(t) $ to an evolution equation of the phase
alone. 
The perturbation ${\bf P}(t)$ is assumed to be sufficiently weak. 
For most practical purposes, the lowest-order perturbation theory
would suffice. Then, a simple method of reduction, which we call
Method I, is applicable.
We begin with a brief presentation
of Method I, and then proceed to a more systematic method including higher order effects of perturbation, called Method II.\par
When the perturbation $\bf{P}$ is introduced, (3.1) is changed to
$$
{\dot\phi}={\rm grad}_{\bf X}\phi\cdot\bigl( {\bf F}({\bf X}) + {\bf P}(t)\bigr)
=\omega+{\rm grad}_{\bf X}\phi\cdot{\bf P}(t). \eqno{(3.2)}
$$
Since our oscillator should still stay near the closed orbit 
$C$, the ${\rm grad}_{\bf X}\phi$ vector may approximately be evaluated
at phase $\phi$ point on $C$ or, equivalently, at the point of intersection 
between $I_\phi$ and $C$. This intersection point is also represented by $\chi(\phi)$. Thus, in what follows, when we say `point 
$\chi (\phi)$', it means the phase $\phi$ point on $C$.\par
Let us use the symbol 
${\bf Z}(\phi)$ for this gradient vector evaluated at $\chi(\phi)$ on $C$, and rewrite
(3.2) as
$$
{\dot\phi}=\omega+{\bf Z}(\phi)\cdot{\bf P}(t). \eqno{(3.3)}
$$
The quantity $\bf Z(\phi)$ is crucial to the study of oscillator dynamics, because it measures
the sensitivity of an oscillator to external stimuli, assumed not too strong,
whatever the kind of the stimuli may be. \par
We have now to specify the kind of perturbation 
${\bf P}(t)$. 
For simplicity, let us first consider a weakly coupled pair of identical oscillators labeled $\alpha$ and $\beta$. 
Then, ${\bf P}(t)$ represents the
coupling force ${\bf V}({\bf X}_\alpha, {\bf X}_\beta)$ acting on oscillator
 $\alpha$.
The mutual coupling is assumed symmetric only for saving suffixes.
The dynamical variables of each oscillator will be specified below by 
the suffixes $\alpha$ or $\beta$, such as ${\bf X}_\alpha$,
${\bf X}_\beta$, $\phi_\alpha$ and $\phi_\beta$. In terms of the coupling 
${\bf V}({\bf X}_\alpha,{\bf X}_\beta)$, we have for oscillator $\alpha$ 
$$
{\dot\phi_\alpha}=\omega+{\bf Z}(\phi_\alpha)\cdot{\bf V}({\bf X}_\alpha, {\bf X}_\beta) \eqno{(3.4)}
$$
and similarly for oscillator $\beta$. In what follows, we do not refer
explicitly to oscillator $\beta$, because the same argument as for 
oscillator $\alpha$ holds simply by interchanging the suffixes $\alpha$
and $\beta$. 
\par
Since we are only concerned with the lowest-order approximation in this subsection, 
${\bf V}({\bf X}_\alpha, {\bf X}_\beta)$ may be replaced with 
${\bf V}(\chi(\phi_\alpha),\chi(\phi_\beta))$. This is a $2\pi$-periodic
 function of either
$\phi_\alpha$ or $\phi_\beta$, and below we abbreviate it as 
${\bf V}(\phi_\alpha,\phi_\beta)$. In this approximation, the phase equation
becomes 
$$
{\dot\phi}_\alpha=\omega+{\bf Z}(\phi_\alpha)\cdot{\bf V}(\phi_\alpha, \phi_\beta). \eqno{(3.5)}
$$
The above equation combined with the similar equation for oscillator 
$\beta$ 
completes the description of the dynamics only in terms of phases. \par
However, this is not the end
of the story. 
To see why, we rewrite (3.5) in terms of the slow 
variables $\psi_{\alpha,\beta}= \phi_{\alpha,\beta} - \omega t$. We have
$$
{\dot \psi}_\alpha={\bf Z}(\omega t+\psi_\alpha)\cdot{\bf V}(\omega t+\psi_\alpha, 
\omega t+\psi_\beta). \eqno{(3.6)}
$$
The above equation is not of a canonical form in the sense that fast oscillation appears parametrically in the evolution equation describing slow dynamics. 
How to remove such fast modulation from the evolution equation constitutes the
second step of reduction. 
This could be achieved by slightly
changing the definition of $\phi_{\alpha,\beta}$.
However, since we are only concerned with the lowest-order approximation, the same results can be obtained more directly by time-averaging~\cite{Kuramoto,ErmentroutKopell,Ermentrout0,Hoppensteadt}. Noting that, on the right-hand side of (3.6) the slow variables $\psi_{\alpha,\beta}$ 
approximately stay constant over one cycle of oscillation of frequency 
$\omega$ because ${\bf V}({\bf X}_\alpha, {\bf X}_\beta)$ is assumed weak, we time-average (3.6) over this period under fixed
$\psi_{\alpha,\beta}$. The resulting equation
takes the simple form
$$
{\dot \psi}_\alpha=\Gamma(\psi_\alpha-\psi_\beta). \eqno{(3.7)}
$$
The effective phase coupling $\Gamma$, which is a $2\pi$-periodic function of the phase difference, is given by
$$
\Gamma(\psi_\alpha-\psi_\beta)=\frac{1}{2\pi}\int^{2\pi}_0 {\bf Z}(y+\psi_\alpha) \cdot
{\bf V}(y+\psi_\alpha, y +\psi_\beta)dy. \eqno{(3.8)}
$$
In terms of the original phases $\phi_{\alpha, \beta}$, (3.7) becomes
$$
{\dot \phi}_\alpha=\omega +\Gamma (\phi_\alpha-\phi_\beta). 
\eqno{(3.9)}
$$
Derivation of the corresponding phase equation for oscillator $\beta$ is trivial.
It is given by ${\dot \phi}_\beta=\omega +\Gamma (\phi_\beta-\phi_\alpha)$.\par
In the above derivation, the two oscillators were assumed identical. However, this condition can be relaxed. Assuming that they are slightly different in nature,
we put ${\bf F}_{\alpha,\beta}({\bf X}_{\alpha,\beta})
={\bf F}({\bf X}_{\alpha,\beta})+\delta{\bf F}_{\alpha,\beta}({\bf X}_{\alpha,\beta})$, where
 $\delta{\bf F}_{\alpha, \beta}({\bf X}_{\alpha,\beta})$ represents a small difference of the 
dynamical system from a suitably defined `average' system represented
by $\bf F(X)$. How to choose this `average' vector field ${\bf F}({\bf X})$
does not affect the final phase equation as long as
 $\delta{\bf F}_{\alpha,\beta}$ remain small. 
Let $\delta{\bf F}_{\alpha}({\bf X}_\alpha)$ be included in ${\bf P}(t)$ of oscillator $\alpha$'s equation
as an additional perturbation. Then, the new term 
$$
\delta\omega_\alpha(\phi_\alpha)\equiv{\bf Z}(\phi_\alpha) \cdot \delta{\bf F}_{\alpha}(\chi(\phi_\alpha)) \eqno{(3.10)}
$$
 must be added to the
right-hand side of (3.5). Similarly to what we did for the coupling function, 
$\delta\omega_\alpha(\phi_\alpha)$ is time-averaged,
resulting in a frequency change by a small constant from $\omega$. \par
The theory developed above can easily be extended so that it may
become applicable to systems of
large numbers of weakly coupled oscillators. The oscillators may be
nonidentical, but only slightly. Asymmetry of mutual coupling is also
allowed.
The result is a popular model of coupled phase oscillators given by~\cite{Kuramoto}
$$
{\dot \phi}_{\alpha}=
\omega_{\alpha}+
\sum_\beta\Gamma_{\alpha\beta}(\phi_{\alpha}-\phi_{\beta}).
\eqno{(3.11)}
$$
See, for example, Refs.~\cite{Winfree0,Kuramoto,Ermentrout0,Ermentrout,Hoppensteadt,Pikovsky,Brown,Ashwin,Nakao0} for the analysis of synchronization in systems of limit-cycle oscillators using the phase model.
\par

\subsection{Method II}

In Method I a number of approximations have been used
in each step of reduction. These approximations are allowed as long as 
we are concerned with the lowest 
order theory. However, by working with this sort of approach,
 we do not see how to generalize the theory
to include systematically higher order effects. Thus, we now come back
to the starting point and move to a more systematic method 
which we call Method II. The theoretical framework of Method II is considerably
different from Method I. Instead, it bears a strong resemblance to
the center-manifold reduction in Sec.~2.\par
The crucial difference from Method I is that we consider explicitly 
a small deviation $\rho$ of the state vector $\bf X$ from the closed orbit
due to the perturbation $\bf P$. Similarly to (2.3), we thus put
$$
{\bf X}(\phi)=\chi (\phi)+\rho(\phi), \eqno{(3.12)}
$$
and let the above equation couple with the phase equation
$$
{\dot\phi}=\omega+G(\phi). \eqno{(3.13)}
$$
The above reduction form is quite similar to (2.4).
Our problem is how to determine the small unknown quantities $\rho$ and $G$ as functions of the phase alone.
Note that the way of splitting the true state vector ${\bf X}(\phi)$ into the reference state $\chi(\phi)$ lying on the closed orbit and the deviation 
$\rho(\phi)$ from this
orbit is generally not unique; we have to fix from what point on 
$C$ the deviation $\rho$ should be measured. However, implicit in (3.12), it has already been assumed that $\bf X$ and $\chi$ have the same phase. That is,
the true state point and the reference state point lie on the same isochron $I_\phi$.
$\rho$ may then be regarded as purely representing amplitude 
disturbance and is completely free from phase disturbance. \par
In order to formulate a systematic perturbation theory, however, curved
hypersurfaces like isochrons will not be an easy object to handle. 
Therefore 
the definition of isophase manifold may desirably be changed slightly. 
In the new definition, the space of equal phase is identified with
the tangent space of $I_\phi$, tangent at point $\chi(\phi)$.
 Let this tangent space be
denoted by $E_\phi$. 
In order that the disturbance vector $\rho$ may completely be free from phase
disturbance, $\rho$ has to be so redefined as to lie 
on $E_\phi$.
Figure~\ref{fig2}(b) shows the isophase surface $E_\phi$ schematically. 
\par
The advantage of working with $E_\phi$ is that it spans 
an eigenspace defined for the linearized system about $C$ or, more precisely, $E_\phi$ is an $(n-1)$-dimensional subspace of this eigenspace. 
Note that for any $\phi$ such an
 eigenspace can be defined. 
The small amplitude disturbance $\rho(\phi)$ may then be decomposed 
into $n-1$ eigenvector components. \par
In a similar way to (2.1), we expand ${\bf F}$, now around $\chi(\phi)$, as
${\bf F}({\bf X}) = {\bf F}(\chi(\phi)) + \epsilon {\hat L}(\phi) \rho(\phi) +{\bf N}(\epsilon \rho(\phi))$, where $\hat{L}(\phi)$ is the Jacobian of ${\bf F}$ evaluated at $\chi(\phi)$ and ${\bf N}(\epsilon \rho)$ represents higher-order terms of $O(\epsilon^2)$.
The eigenspace and eigenvectors are 
defined for the
linearized equation 
$$
{\dot \rho}={\hat L}(\omega t)\rho \eqno{(3.14)}
$$
about the limit-cycle solution $\chi(\omega t)$.
Since the Jacobian ${\hat L}$ is time-periodic, using the Floquet theorem~\cite{GuckenheimerHolmes},
the solution $\rho$ can be written in the form
$\rho(\phi) = \hat{S}(\phi) e^{(\phi / \omega) \hat\Lambda(0)} \rho(0)$ or, more generally,
$\rho(\phi + \psi) = \hat{S}(\phi+\psi) e^{(\psi / \omega) \hat\Lambda(0)} \hat{S}^{-1}(\phi) \rho(\phi)$,
where ${\hat S}(\phi)$ satisfies ${\hat S}(\phi+2\pi)={\hat S}(\phi)$ and $\hat{S}(0)=I$.
If we make a stroboscopic 
observation of $\rho$ at time interval of the period $T=2\pi/\omega$,
we obtain $\rho(\phi + 2\pi) = \hat{S}(\phi) e^{ T \hat{\Lambda}(0) } \hat{S}^{-1}(\phi) \rho(\phi) = e^{ T \hat{\Lambda}(\phi) } \rho(\phi)$, where $\hat\Lambda(\phi)$ is a time-independent but $\phi$-dependent $n\times n$ matrix
given by $\hat\Lambda(\phi) = \hat{S}(\phi) \hat\Lambda(0) \hat{S}^{-1}(\phi)$.
Thus, the above linear equation is transformed to an ordinary eigenvalue problem 
$e^{T \hat\Lambda(\phi)}\rho(\phi)=e^{\lambda T} \rho(\phi)$, or
${\hat\Lambda}(\phi) \rho(\phi) = \lambda \rho (\phi)$, where $\lambda$ is the eigenvalue called the Floquet exponent.
There are $n$ eigenvalues, $\lambda_{0, ..., n-1}$. Among them, one eigenvalue is zero and denoted as $\lambda_0 = 0$. All other eigenvalues have negative real parts, $\mbox{Re} \lambda_{1, ..., n-1} < 0$, because the limit cycle is stable.
\par
Let the eigenvectors of ${\hat \Lambda}(\phi)$ be denoted by  
${\bf u}_{l}(\phi)$
$(l=1,2,\ldots,n)$. Out of these eigenvectors, $n-1$ eigenvectors 
${\bf u}_1(\phi),{\bf u}_2(\phi),\ldots,{\bf u}_{n-1}(\phi)$ with
nonzero eigenvalues span $E_\phi$.
The remaining eigenvector 
${\bf u}_0(\phi)$
with zero eigenvalue corresponds to infinitesimal phase disturbance.
The direction of ${\bf u}_0(\phi)$ is tangential to the limit-cycle orbit 
$C$ at $\chi(\phi)$. Thus, this eigenvector is parallel
with $d\chi(\phi)/d\phi$. In what follows, as a convenient choice, 
${\bf u}_0(\phi)$
will be identified with $d\chi(\phi)/d\phi$.
\par
One may also define left-eigenvectors of $\hat\Lambda(\phi)$, which we denote by ${\bf u}^\ast_0(\phi),$
${\bf u}^\ast_1(\phi)$, ${\bf u}^\ast_2(\phi)$, $\ldots$, ${\bf u}^\ast_{n-1}(\phi)$
satisfying biorthonormal relations with the right eigenvectors as ${\bf u}^\ast_l(\phi) {\bf u}_m(\phi) = \delta_{l,m}$}.
It should be noted that the sensitivity vector ${\bf Z}(\phi)$ is 
identical with ${\bf u}^\ast_0(\phi)^{\dag}$, where $\dag$ denotes transpose. 
The reason is that 
the orthonormal condition required for ${\bf u}_0^\ast(\phi)^{\dag}$ is also satisfied by 
${\bf Z}(\phi)$. Regarding the orthogonality condition, remember that
the phase gradient vector ${\rm grad}_{\bf X}\phi$ must be vertical
to the isophase hypersurface. This means that ${\bf Z}(\phi)$ is vertical to
$E_{\phi}$, or ${\bf Z}(\phi) \cdot {\bf u}_l(\phi)=0$ $(l\ne 0)$. Normalization
condition is clearly satisfied because 
${\bf Z}(\phi) \cdot {\bf u}_0(\phi)
=({\rm grad}_{\bf X}\phi)_{{\bf X
}=\chi(\phi)} \cdot (d{\bf X}/d\phi)_{{\bf X}=\chi(\phi)}=1$.
 \par
Another technical remark should be made. 
As stated above, the 
eigenspace is defined at each phase on $C$. There is a
relationship between the eigenvectors belonging to the same 
eigenvalue at
different $\phi$. By choosing the zero phase $\phi=0$ as 
the reference phase, ${\bf u}_l(\phi)$ is given by a transformation
of ${\bf u}_l(0)$ with the matrix ${\hat S}(\phi)$ common to all $l$:
$$
{\bf u}_l (\phi)={\hat S}(\phi){\bf u}_l (0). \eqno{(3.15)}
$$
Clearly, there is also a corresponding relation for 
the left-eigenvectors:
$$
{\bf u}_l^\ast (\phi)={\bf u}_l^\ast (0){\hat S}^{-1}(\phi). \eqno{(3.16)}
$$
It can also be shown by differentiating the solution $\rho(\phi)$ that the Jacobian ${\hat L}(\phi)$ is related to ${\hat \Lambda}(0)$ through 
$$
{\hat \Lambda}(0)={\hat S}^{-1}(\phi)\Bigl(L(\phi)-\omega\frac{d}{d\phi} \Bigr){\hat S}(\phi). \eqno{(3.17)}
$$
\par
In the past, Method II was not formulated for multi-oscillator systems except for oscillatory reaction-diffusion systems; space-discrete
systems such as networks and populations of oscillators were
carefully avoided~\cite{Kuramoto}. In discrete systems,
the pair coupling, which has to be treated as a perturbation,
depends on the two state vectors of the coupled pair like 
$\bf V(\bf X, \bf X')$. This fact seemed to make it difficult
to treat the problem formally as a one-oscillator problem. 
As we see later in this section, this seeming difficulty can be circumvented easily. However, before going into this problem, we begin with an easier case to treat, because the formulation is almost the same as for discrete systems except for one essential point.
\par
The following example may be practically uninteresting but seems quite instructive for illustrating the general structure of the theory. 
Suppose that the perturbation ${\bf P}(t)$ is simply given by a function only of 
$\bf X$ of the oscillator of concern. This means that the nature of
the oscillator has been slightly changed. By putting ${\bf P}(t)=\epsilon \delta{\bf F}(\bf X)$, where $\delta{\bf F}({\bf X})$ represents the change in the oscillator dynamics, the equation to be reduced is given by
$$
{\dot {\bf X}}={\bf F(X)}+\epsilon \delta{\bf F}(\bf X). \eqno{(3.18)}
$$ 
We have introduced parameter $\epsilon$ as an indicator of the smallness of the perturbation but we will put $\epsilon=1$ at the end of the calculation. Our goal is
to find the solution of the problem in an $\epsilon$-expansion form. 
\par
As stated repeatedly, we try to solve the reduction problem in 
the following form:
$$
 {\bf X}(t) =\chi (\phi)+\epsilon\rho(\phi), \eqno{(3.19)}
$$
$$
{\dot \phi}=\omega+\epsilon G(\phi). \eqno{(3.20)}
$$
In the above equation, $\rho(\phi)$ and $G(\phi)$ generally include 
higher order corrections and can still be expanded in powers of 
$\epsilon$.
\par
To determine the functional form of the unknowns $\rho(\phi)$ and $G(\phi)$, 
we insert (3.19) into (3.18), and use (3.20) to replace the time-derivative
with $\phi$-derivative.
With the use of the identity (3.17),
the result is concisely summarized in the form 
$$
G(\phi){\bf u}_0 (0)-\Bigl({\hat \Lambda}(0)-\omega\frac{d}{d\phi} \Bigr)
{\hat S}^{-1}(\phi)\rho(\phi)={\hat S}^{-1}(\phi){\bf B}(\phi). 
\eqno{(3.21)}
$$
The $\bf B$ term consists primarily of the lowest-order perturbation
${\delta\bf F}(\chi(\phi))$ but also includes smaller quantities
depending on $\rho$ and $G$. That is, 
$$
{\bf B}(\phi)=\delta{\bf F}\bigl(\chi(\phi)+ \epsilon \rho(\phi) \bigr) 
- \epsilon G(\phi)\frac{d \rho(\phi)}{d \phi}
+ \epsilon^{-1} {\bf N}(\epsilon \rho(\phi)). \eqno{(3.22)}
$$
There is a strong similarity of (3.21) to the corresponding equation
(2.5) in the center-manifold reduction. The only important difference
between these equations is
that the reduced phase equation or $G(\phi)$ to be obtained from 
(3.21) does not have a canonical form yet, while in the center-manifold 
reduction the canonical form was directly obtained by solving (2.5).
\par
Note that $G(\phi)$ appears as the coefficient of the 
zero eigenvector ${\bf u}_0(0)$ of ${\hat \Lambda}(0)$. 
Note also that, by decomposing $\rho(\phi)$ as
$\rho(\phi)=\sum_{l\ne 0}c_l (\phi){\bf u}_l(\phi)$, we have
$$
{\hat S}^{-1}(\phi)\rho (\phi)=\sum_{l\ne 0}c_l(\phi){\hat S}^{-1}(\phi){\bf u}_l(\phi)
=\sum_{l\ne 0} c_l(\phi){\bf u}_l (0). \eqno{(3.23)}
$$
Thus, if we pretend that $\bf B$ is a known quantity, then (3.21) represents a set of linear uncoupled equations for the $\phi$-dependent coefficients of the eigenvectors ${\bf u}_l(0)$. The coefficient of
 ${\bf u}_0(0)$ is given by
$G(\phi)$ which is the only quantity of our ultimate concern.
Each coefficient will be found by taking the scalar product of the left
eigenvector ${\bf u}_l^\ast(0)$ with each side of (3.21).
In particular, by taking a scalar product with ${\bf u}_0^*(0)$,
$G(\phi)$ is given by 
$$
G(\phi)={\bf u}_0^{\ast}(0){\hat S}^{-1}(\phi){\bf B}(\phi)=
{\bf u}_0^{\ast}(\phi){\bf B}(\phi). \eqno{(3.24)}
$$
Each of the other coefficients $c_l(\phi)$ is given by the solution of the first order differential equation
$$
\Bigl(\lambda_l -\omega\frac{d}{d\phi} \Bigr)c_l(\phi)
=-{\bf u}_l^\ast(\phi){\bf B}(\phi). \eqno{(3.25)}
$$
The above equation is solved by integration, whose solution $c_l(\phi)$ must be a 
periodic function. Noting that the real part of the eigenvalue $\lambda_l$ is negative,
one may easily show that the periodicity condition can be satisfied by taking the 
lower limit of the integral to be $-\infty$.
\par
By starting with the $\bf B$ in the lowest-order approximation, which is given by ${\bf B}(\phi)=\delta{\bf F}(\chi(\phi))$, the unknowns $\rho(\phi)$ and $G(\phi)$ will be found iteratively in the form of $\epsilon$-expansion. In particular, the lowest-order $G$ becomes
$G(\phi)={\bf u}^\ast_0(\phi)\delta{\bf F}(\chi(\phi))$, and this coincides with the result from Method I given by (3.10). How to solve the problem to any desired order
in $\epsilon$ by iteration will no longer need explanation. 
\par
We are now ready to deal with populations of weakly coupled oscillators. 
In order to see the source of the aforementioned difficulty about such systems and how this difficulty is circumvented, the study of the following simple system would be sufficient. This is a pair of identical oscillators with weak symmetric coupling, which was also studied in Method I.
The governing 
equation for oscillator $\alpha$ is given by 
$$
{\dot {\bf X}}_\alpha={\bf F}({\bf X}_\alpha)+\epsilon{\bf V}({\bf X}_\alpha,{\bf X}_\beta) \eqno{(3.26)}
$$
and similarly for oscillator $\beta$.
\par
As a generalization of (3.19) and (3.20), it seems natural to assume the 
reduction form as
$$
 {\bf X}_\alpha(t) =\chi (\phi_\alpha)+\epsilon\rho(\phi_\alpha,\phi_\beta), \eqno{(3.27)}
$$
$$
{\dot \phi}_\alpha=\omega+\epsilon G(\phi_\alpha,\phi_\beta). \eqno{(3.28)}
$$
A similar reduction form is also assumed for oscillator $\beta$. The above form must satisfy (3.26). This requirement leads to the following equation similar to (3.21):
$$
G(\phi_\alpha,\phi_\beta){\bf u}_0(0)
-\bigl({\hat \Lambda}(0)-\omega\frac{d}{d\phi_\alpha} \bigr)
{\hat S}^{-1}(\phi_\alpha)\rho(\phi_\alpha,\phi_\beta)={\hat S}^{-1}(\phi_\alpha){\bf B}(\phi_\alpha,\phi_\beta). \eqno{(3.29)}
$$
The problem arising here is that the lowest-order $\bf B$ contains an unknown 
quantity $\rho$. In fact, we have
$$
{\bf B}(\phi_\alpha,\phi_\beta)={\bf V}\bigl(\chi(\phi_\alpha),\chi(\phi_\beta)\bigr)
-\omega\frac{\partial\rho(\phi_\alpha,\phi_\beta)}{\partial\phi_\beta}
+O(\epsilon) {\ \rm terms}, \eqno{(3.30)}
$$
where the $\partial\rho/\partial\phi_\beta$ term
is comparable with the $\bf V$ term in magnitude. 
Thus, finding the solution by successive iterations seems impossible.
Still, $G$ in the lowest order can be determined, because
the unwanted quantity $\rho$, and hence the $\partial\rho/\partial\phi_\beta$ term, contained in $\bf B$, is free from the
${\bf u}_0(0)$ component. \par
As stated before, in phase reduction and also in center-manifold reduction,
the problem should be treated formally as a one-oscillator problem
even if we are dealing with many-oscillator systems.
In the present case, this means that no quickly changing variable other 
than $\phi_\alpha$ must appear in the formulation. If this is violated, then 
the time derivatives of $\rho$ through such fast variables can not be small
and contribute to the lowest-order terms in $\bf B$, thus making iteratively 
solving the problem unfeasible from the outset. \par
Actually, how to circumvent this difficulty is simple. 
We only need to change
the independent variables
from $\phi_\alpha$ and $\phi_\beta$ to $\phi_\alpha$ and $\psi\equiv \phi_\alpha-\phi_\beta$. With the use of simplified notations $\rho(\phi_\alpha,\psi)$, $G(\phi_\alpha,\psi),$$\ldots$ in place of $\rho(\phi_\alpha,
\phi_\alpha-\psi)$, $G(\phi_\alpha, \phi_\alpha-\psi)$,$\ldots$, we have now to find the solution in the form
$$
 {\bf X}_\alpha(t) =\chi (\phi_\alpha)+\epsilon\rho(\phi_\alpha,\psi),
\eqno{(3.31)}
$$
$$
{\dot \phi}_\alpha=\omega+\epsilon G(\phi_\alpha,\psi).
\eqno{(3.32)}
$$
By interchanging the suffixes $\alpha$ and $\beta$ and 
replacing $\psi$ with $-\psi$, we have the corresponding equations for 
oscillator $\beta$. As a result,
in place of (3.29), we have
$$
G(\phi_\alpha,\psi){\bf u}_0(0)
-\Bigl({\hat \Lambda}(0)-\omega\frac{d}{d\phi_\alpha} \Bigr)
{\hat S}^{-1}(\phi_\alpha)\rho(\phi_\alpha,\psi)={\hat S}^{-1}(\phi_\alpha){\bf B}(\phi_\alpha,\psi). \eqno{(3.33)}
$$
No unpleasant term appears in $\bf B$ now, and the lowest-order $\bf B$ is simply given by
$\bf B(\phi_\alpha, \psi)={\bf V}(\chi(\alpha),\chi(\phi_\alpha-\psi))$. Although $\bf B$ still includes the time derivative 
of $\rho$ through the second variable $\psi$, this gives a small quantity
because $\dot\psi=
\epsilon(G(\phi_\alpha,\psi)-G(\phi_\alpha-\psi, -\psi))$. \par
In this way, $G(\phi_\alpha,\psi)$ and $\rho(\phi_\alpha,\psi)$ can be 
determined by successive iterations to any desired order in $\epsilon$,
although all calculational details are omitted here. Coming back to the 
representation in terms of the original phase variables $\phi_\alpha$
and $\phi_\beta$, we arrive at the 
$\epsilon$-expansion form
$$
{\dot \phi}_\alpha=\omega+\epsilon G_1(\phi_\alpha,\phi_\beta)
+\epsilon^2 G_2(\phi_\alpha,\phi_\beta)+\ldots. \eqno{(3.34)}
$$
Note that the above idea of treating the phase difference as a new
independent variable is similar to the past study in which we applied Method II
to diffusion-coupled systems where spatial derivatives of the phase
were regarded as independent variables~\cite{Kuramoto}.\par
Although the calculation becomes even more involved, there is no 
essential difficulty in extending the above-developed method to
a large assembly of weakly coupled oscillators. The only point to be
noticed is that, since no fast variables
other than $\phi_\alpha$ must appear, we have to introduce many 
$\psi$-variables as $\psi_\beta=\phi_\alpha-\phi_\beta,\:\psi_\gamma=\phi_\alpha-\phi_\gamma,\: \ldots$. The resulting phase equation 
in terms of the original phases takes 
the form
$$
{\dot \phi}_\alpha=\omega+\epsilon\sum_\beta G_1(\phi_\alpha,\phi_\beta)
+\epsilon^2 \Bigl(\sum_\beta G_2^{(1)}(\phi_\alpha,\phi_\beta)+
\sum_{\beta,\gamma}G_2^{(2)}(\phi_\alpha,\phi_\beta,\phi_\gamma)+\cdots
\Bigr) . \eqno{(3.35)}
$$
As a nontrivial feature of the above equation, many-body coupling 
appears in the second and higher order terms. One of the $\epsilon^2$-order terms depends on 3 variables. Generally, the maximum number
of phase variables appearing in the coupling terms increases one by one with the increasing power of $\epsilon$. 
\par
We now move to the second step of reduction where we transform the phase equation to a canonical form. As noted in Method I, the canonical form means that the coupling function depends only on 
the phase difference. To illustrate how this is achieved, let us work with
the lowest-order phase equation for a pair of symmetrically coupled 
oscillators. Thus, we consider the equation
$$
{\dot \phi}_\alpha=\omega+\epsilon G(\phi_\alpha,\phi_\beta)
\eqno{(3.36)}
$$
coupled with the equation for oscillator $\beta$, i.e.
${\dot\phi}_\beta=\omega+\epsilon G(\phi_\beta, \phi_\alpha)$.
The essentials of the theory could fully be
explained with this example. \par
To transform the above coupled phase equations to a desired form, we require a
near-identity transformation from $\phi_\alpha$ and $\phi_\beta$ to new variables ${\tilde \phi}_\alpha$ and
 ${\tilde \phi}_\beta$. The new variables will be so chosen that the coupling may become a function of the phase difference alone. 
Thus, in terms of the phase difference $\psi\equiv {\tilde \phi}_\alpha
-{\tilde \phi}_\beta$, we require for oscillator $\alpha$ the following reduction form similar to (3.31) and (3.32):
$$
\phi_\alpha={\tilde \phi}_\alpha+\epsilon \sigma({\tilde\phi}_\alpha, \psi),
 \eqno{(3.37)}
$$
$$
{\dot {\tilde\phi}}_\alpha=\omega+\Gamma(\psi). \eqno{(3.38)}
$$
To make the above transformation unique, we impose the boundary condition
$\sigma (0,\psi)=0$ which means that ${\tilde \phi}_\alpha=0$ implies
$\phi_\alpha=0$. 
As in the first step of reduction, working with $\phi_\alpha$ and $\psi$
and not with ${\tilde\phi}_\alpha$ and ${\tilde\phi}_\beta$ is crucial. \par
The reduction form (3.37) and (3.38) must satisfy (3.36). 
The result of substituting (3.37) and (3.38) into (3.36) may be
summarized as
$$
\frac{\partial\sigma({\tilde\phi}_\alpha)}{\partial{\tilde\phi}_\alpha}
=\omega^{-1} \bigl(-\Gamma(\psi)+B({\tilde\phi}_\alpha,\psi) \bigr),
\eqno{(3.39)}
$$
$$
B({\tilde\phi}_\alpha,\psi) = G({\tilde\phi}_\alpha,{\tilde\phi}_\alpha-\psi)+O(\epsilon)
 {\ \rm terms}. \eqno{(3.40)}
$$
The quantity $B$ consists primarily of the coupling term in the lowest-order
approximation but also contains some other terms depending on the unknown
quantities. The situation is similar to the first step, that is, 
if $\sigma$ and $\Gamma(\psi)$ are determined
simultaneously from (3.39) with $B$ supposed known, then they must be found to any
desired precision by iteration. 
This is in fact possible. Note that $\sigma$ is a $2\pi$-periodic function 
of ${\tilde \phi}_\alpha$, which means that it vanishes by integration over
one period. This condition determines $\Gamma$ as
$$
\Gamma(\psi)=\frac{1}{2\pi}\int^{2\pi}_0 B(y,\psi)dy\equiv B_{av}(\psi).
\eqno{(3.41)}
$$
Then, $\sigma({\tilde \phi}_\alpha,\psi)$ is obtained by integrating
the right-hand side of (3.39):
$$
\sigma({\tilde\phi}_\alpha,\psi)=\omega^{-1}\int^{{\tilde\phi}_\alpha}_0
\bigl(B(y,\psi)-B_{av}(\psi)\bigr)dy. \eqno{(3.42)}
$$
It is clear that, starting with the lowest-order $ B$ obtained by dropping the $O(\epsilon)$ terms in (3.40), we can achieve successive approximations for $\sigma$ and $\Gamma$.
\par
This completes the second step of reduction, but only for the simple phase equation (3.36). There are at least four conditions to be relaxed
by which the formulation of the second step of reduction can be generalized. First, higher order terms may be included in (3.36) with 
which we start. As a result, many-body coupling will necessarily be
included. Second, larger assembly of oscillators
than an oscillator pair can be treated. Third, the coupling may be
asymmetric. Fourth, the oscillators may be slightly non-identical.
The phase equation including all these facts will take the form
$$
{\dot \phi}_\alpha=\omega_\alpha(\phi_\alpha)+\epsilon\sum_\beta
G_{\alpha,\beta,1}(\phi_\alpha,\phi_\beta) 
+\epsilon^2\Bigl(
\sum_\beta G_{\alpha,\beta,2}^{(1)}(\phi_\alpha,\phi_\beta)+
\sum_{\beta,\gamma}G_{\alpha,\beta,\gamma,2}^{(2)}(\phi_\alpha,\phi_\beta,\phi_\gamma) \Bigr)+\ldots \eqno{(3.43)}
$$
To integrate all such complex situations into the theory
would not be a difficult matter if we do not mind lengthy calculations.
Omitting all
calculational details, we arrive at the final form of phase reduction
$$
{\dot\phi}_\alpha=\omega_\alpha+\epsilon\sum_\beta \Gamma_1(\phi_\alpha-\phi_\beta)+\epsilon^2\Bigl(
\sum_\beta \Gamma_2^{(1)}(\phi_\alpha-\phi_\beta)+
\sum_{\beta,\gamma}G_2^{(2)}(\phi_\alpha-\phi_\beta,\phi_\alpha-\phi_\gamma) \Bigr)+\ldots. \eqno{(3.44)}
$$

Finally, one may wonder if there is any practical use of the theory of
higher order phase reduction. It seems difficult to give a
definite answer as yet, 
but it may at least be said that
the higher order corrections will become important when the lowest-order
phase equation breaks down. For the phase
reduction of reaction-diffusion system~\cite{Kuramoto}, this is actually the case 
because the lowest-order equation of the Burgers type breaks down when 
the phase diffusion constant $\nu$ becomes negative. As long as $\nu$ 
remains small negative, the next order theory 
which produces the fourth order spatial derivative of phase
becomes quite meaningful. Although not confirmed yet, 
similar situation could be met in discrete assemblies of oscillators as well.
Imagine either a short-range coupled oscillator lattice or all-to-all 
coupled population. It may happen that the lowest-order 
coupling function stays near the threshold between the in-phase and
anti-phase types. The resulting dynamical state, whatever it may be,
will have a weak stability, and some additional perturbation may easily cause a
drastic change in the dynamics. It seems reasonable to expect
that the higher order terms in the phase equation will play the role of 
such an extra perturbation. 

\section{Concluding remarks}

We have seen how the two representative theories of reduction
are similar to each other in their basic way of thinking as well as their
theoretical structure.
Each theory must involve two essential steps, namely,
reduction of dynamical degrees of freedom on one hand and 
transformation of the reduced equation to a canonical form on the 
other hand. In the center-manifold reduction, however, these steps
need not be taken one by one but could be achieved simultaneously.
This was possible because the complete set of eigenfunctions of the
operator ${\hat L}_0-\omega d/d\theta$ is simply given by the product of
the eigenvectors of ${\hat L}_0$ and the various harmonics $e^{im\theta}$,
thus making the coefficients of these eigenfunctions become 
independent of $\theta$. Such a nice property cannot be expected 
for the phase reduction, because this theory, unlike the center-manifold 
reduction, deals with periodic orbits without high symmetry. \par
Another important feature common to the two reduction theories is that
even when we deal with multi-oscillator systems the theory is formulated
formally as a perturbed one-oscillator problem. Thus, when the center-manifold
reduction is applied to reaction-diffusion systems, for example, 
we still imagine a two-dimensional center manifold depending on various spatial derivatives of the amplitude $A$ as parameters.\par
Finally, a few remarks will be given on the significance of dynamical reduction
from a broader perspective of science in general. There seem to be
several roles to be played by the reduction theories for our 
understanding, controlling and designing systems around us. The most primitive motivation
of the reduction theories was that, when we were given a complex dynamical system model, we expected that its analysis would become far easier
by reduction. However, it is important to notice that reduced equations are derived without knowledge of the specific form of the original evolution equation.
With the help of reduction theories, we often understand the reason for this or that observed behavior of complex systems for which we are unable 
to construct a dynamical system model. 
\par
Predictive power of reduction is also great. Indeed, it sometimes happens that unexpectedly new types of dynamics are discovered through the analysis of 
reduced equations. Such dynamical behavior is expected to exist universally
because the reduced equations themselves are universal. For example, 
we owe much to the complex Ginzburg-Landau equation and its variants
for our far deeper understanding of continuous fields of oscillators 
than before. Discovery of the concept of phase turbulence is also a fruit
of dynamical reduction~\cite{Kuramoto}.\par
Implicit in the above arguments, we are thinking of qualitative understanding
rather than quantitative understanding of the nature. Although quantitative 
validity of the reduced equations is limited to the vicinity of exceptional 
situations such as the near-critical situation of bifurcation and vanishingly small coupling strength, such restriction is not so severe for qualitative
understanding. Whenever
necessary, one may come back to more realistic mathematical models.
Their analysis will become far easier with the help of the suggestions
provided by the analysis of the reduced equations. Furthermore,
regarding how to control coupled oscillator systems as desired, 
 there may be a lot to learn from the analysis of reduced equations. 
This seems true whether the coupled oscillators concerned exist 
as a natural system or they are designed as a man-made system.

\section{Appendix}

In this appendix, we briefly describe some of the recent developments in the phase-reduction approach to limit-cycle oscillators, that is, the phase-amplitude reduction, Koopman-operator viewpoint, and some extensions of the phase reduction to non-conventional dynamical systems.

\subsection{Phase-amplitude reduction}

In Sec.~3, we formulated the phase reduction for weakly perturbed limit-cycle oscillators. As we explained, the phase is the only slow variable describing the long-time asymptotic dynamics of the oscillator. The amplitude disturbance, representing deviation of the oscillator state from the limit cycle, was neglected in Method I or eliminated in Method II.
However, when we are interested in the transient dynamics of the oscillator at shorter timescales, such as the relaxation dynamics to the limit cycle or small variation from the limit cycle caused by applied perturbations, the amplitude degrees of freedom may also be important.

Recently, phase-amplitude reduction of the oscillator dynamics, which explicitly takes the amplitude degrees of freedom into account, has been formulated~\cite{Wedgwood,Wilson,Mauroy1,Shirasaka}. These studies are partly inspired by the recent development of the Koopman-operator approach to nonlinear dynamical systems, which we briefly discuss in the next subsection, though the idea of the phase-amplitude description of the oscillator state around a limit cycle had already been discussed, for example, in the classical textbook of differential equations by  Hale~\cite{Hale}.
In this subsection, we briefly describe the extension of Method I of phase reduction to phase-amplitude reduction.

First, we need to introduce the amplitude variables that describe deviations of the oscillator state from the limit cycle $C$.
In Sec.~3, the phase $\phi({\bf X})$ of the oscillator state ${\bf X}$ was defined to satisfy Eq.~(3.1), i.e. ${\dot\phi}={\bf F(X)} \cdot {\rm grad}_{\bf X}\phi =\omega$, where $\omega$ is the frequency of the oscillator.
Considering that small deviation of the unperturbed oscillator state from $C$ decays exponentially on average, it seems natural and convenient to introduce an amplitude function $r({\bf X})$ which satisfies 
$$
\dot{r} = {\bf F(X)} \cdot {\rm grad}_{\bf X} r = \lambda r
\eqno{(5.1)}
$$
in the basin of $C$, where $\lambda$ is the decay rate.
The amplitude $r$ vanishes on the limit cycle, i.e. $r(\chi(\phi)) = 0$ where $\chi(\phi)$ is the point of phase $\phi$ on $C$.
From the Floquet theory, it is expected that $\lambda$ should be equal to one of the Floquet exponents $\lambda_{1, ..., n-1}$ with negative real parts (sorted as $\lambda_0 = 0 > \mbox{Re} \lambda_1 \geq \mbox{Re} \lambda_2 \cdots \geq \mbox{Re} \lambda_{n-1}$)
and that $n-1$ such amplitude variables can be introduced. This is actually the case as explained below.
For simplicity, we assume that the Floquet exponents are simple and real; extension to the complex case is straightforward.

Suppose that such amplitude functions $r_l({\bf X})$ ($l=1, ..., n-1$) exist. Then, as in Sec. 3, the amplitude $r_l$ of a weakly perturbed oscillator $\dot{\bf X} = {\bf F}({\bf X}) + {\bf P}(t)$ obeys
$$
{\dot r}_l = {\rm grad}_{\bf X} r_l \cdot \bigl( {\bf F}({\bf X}) + {\bf P}(t) \bigr)
= \lambda_l r_l + {\rm grad}_{\bf X} r_l \cdot {\bf P}(t)
\eqno{(5.2)}
$$
for each amplitude variable.
Moreover, if ${\bf P}(t)$ is sufficiently weak and ${\bf X}$ stays in the close vicinity of $C$, the gradient vector of $r_l({\bf X})$ may be approximately evaluated on $C$ in the lowest-order approximation.  We thus obtain an approximate amplitude equation
$$
{\dot r}_l = \lambda_l r_l + {\bf I}_l(\phi) \cdot {\bf P}(t)
\eqno{(5.3)}
$$
for $l=1, ..., n-1$, where ${\bf I}_l(\phi) = {\rm grad}_{\bf X} r_l|_{{\bf X} = \chi(\phi)}$ is the gradient vector of $r_l$ evaluated at $\chi(\phi)$ on $C$ characterizing the sensitivity of $r_l$ to external stimuli. It is important to note that ${\bf I}_l(\phi)$ depends only on the phase and not on the amplitudes, because it is evaluated on the limit cycle $C$. Note the similarity of Eq.~(5.3) and the function ${\bf I}_l(\phi)$ for the amplitude to Eq.~(3.3) and the sensitivity function ${\bf Z}(\phi)$ for the phase.
In Sec. 3, it was shown that the sensitivity function ${\bf Z}(\phi)$ of the phase is equal to the transpose of the left eigenvector ${\bf u}_0^*(\phi)$ of $\hat{\Lambda}(\phi)$ associated with the zero eigenvalue.
Similarly, ${\bf I}_l(\phi)$ can be chosen as the $l$th left eigenvector of $\hat{\Lambda}(\phi)$ as ${\bf I}_l(\phi) = {\bf u}_l^*(\phi)^{\dag}$ as shown below.

Let us now derive the above facts, namely, the decay rate $\lambda$ of $r$ in Eq.~(5.1) is equal to one of the Floquet exponents $\lambda_{1, ..., n-1}$ and ${\bf I}_l(\phi) = {\bf u}_l^*(\phi)^{\dag}$ ($l=1, ..., n-1$).
Suppose that an initial oscillator state at $t=0$ is chosen as ${\bf X}(0) = \chi(0) + \epsilon \rho(0)$ with $\phi=0$, where $\rho(0) = {\bf u}_j(0)$ ($j=0, ..., n-1$) and $\epsilon$ is an arbitrary tiny parameter; namely, we add a tiny amplitude disturbance to the oscillator state in the direction of $j$th eigenvector of $\hat{\Lambda}(0)$ with eigenvalue $\lambda_l$.
Evolving the oscillator without perturbation from this initial state under the linear approximation, the oscillator state at $t = \phi / \omega$ can be expressed as ${\bf X}(t) = \chi(\phi) + \epsilon \rho(\phi) = \chi(\phi) + \epsilon \hat{S}(\phi) e^{( \phi / \omega ) \hat{\Lambda}(0)} {\bf u}_j(0) = \chi(\phi) + \epsilon \hat{S}(\phi) e^{\lambda_j ( \phi / \omega ) } {\bf u}_j(0)$ as in Sec. 3.
Expanding $r({\bf X})$ in Taylor series as $r(\chi(\phi) + \epsilon \rho(\phi)) = r(\chi(\phi)) + {\rm grad}_{\bf X} r|_{{\bf X} = \chi(\phi)} \cdot \epsilon \rho(\phi) + O(\epsilon^2) = \epsilon {\bf I}(\phi) \cdot \rho(\phi) + O(\epsilon^2)$, where ${\bf I}(\phi) = {\rm grad}_{\bf X} r|_{{\bf X} = \chi(\phi)}$ is the gradient of $r$ on $C$, the amplitude $r$ of ${\bf X}(t)$ can be expressed as $r({\bf X}(t)) = \epsilon {\bf I}(\phi) \cdot \rho(\phi) + O(\epsilon^2) = \epsilon {\bf I}(\phi) \cdot \hat{S}(\phi) e^{ \lambda_j ( \phi / \omega )} {\bf u}_j(0) + O(\epsilon^2)$.
On the other hand, the amplitude $r$ should obey Eq.~(5.1) by assumption and thus $r({\bf X}(t)) = e^{\lambda t} r({\bf X}(0)) = \epsilon e^{\lambda t} {\bf I}(0) \cdot {\bf u}_j(0)$.
Therefore, ${\bf I}(\phi) \cdot \hat{S}(\phi) e^{\lambda_j ( \phi / \omega ) } {\bf u}_j(0) = e^{\lambda (\phi / \omega)} {\bf I}(0) \cdot {\bf u}_j(0)$ should hold for any $\phi$ within the linear approximation.
This requires the following two conditions to be satisfied: (i) $ {\bf I}(0) \cdot {\bf u}_j(0) = 0$ if $\lambda \neq \lambda_j$, and (ii) ${\bf I}(\phi) \cdot \hat{S}(\phi) = {\bf I}(0)$ if $\lambda = \lambda_j$.
From the condition (i), it turns out that a nontrivial amplitude $r$ can only be defined when $\lambda$ is equal to any of $\lambda_{1, ..., n-1}$, otherwise $r$ takes constantly $0$ for any $\rho(0)$ within the linear approximation. 
We thus assume $\lambda = \lambda_l$ and denote the corresponding amplitude and gradient as $r_l$ and ${\bf I}_l$, respectively.
Then, if $l \neq j$, $ {\bf I}_l(0) \cdot {\bf u}_j(0)$ should be $0$, and if $l = j$, ${\bf I}_l(0) \cdot {\bf u}_l(0)$ should not be $0$.
These conditions can be satisfied only if we take ${\bf I}_l(0) \propto {\bf u}^*_l(0)^{\dag}$ for $l=1, ..., n-1$, and we simply adopt ${\bf I}_l(0) = {\bf u}^*_l(0)^{\dag}$.
We then obtain ${\bf I}_l(\phi)^{\dag} = {\bf I}_l(0)^{\dag} \hat{S}^{-1}(\phi)$, i.e. ${\bf I}_l(\phi) = {\bf u}_l^*(\phi)^{\dag}$ from the condition (ii).

It is also important that these left eigenvectors, or sensitivity functions, can be numerically evaluated much more easily than the amplitude function itself. For the sensitivity function of the phase, it is well known that ${\bf Z}(\phi) = {\bf u}^*_0(\phi)^\dag$ is a $2\pi$-periodic solution to an adjoint linear equation $\omega d{\bf Z}(\phi)/d\phi = - \hat{L}(\phi)^{\dag} {\bf Z}(\phi)$ with normalization condition ${\bf Z}(\phi) \cdot {\bf F}(\chi(\phi)) = \omega$~\cite{Ermentrout}. See 
also Ref.~\cite{Brown} for a very concise derivation and the functional forms of ${\bf Z}(\phi)$ for typical bifurcations of limit cycles.
Using a similar argument, it can be shown that ${\bf I}_l(\phi) = {\bf u}_l^*(\phi)^\dag$ is a $2\pi$-periodic solution to an adjoint linear equation, $\omega d{\bf I}_l(\phi)/d\phi = - [ \hat{L}(\phi)^{\dag} - \lambda_l ] {\bf I}_l(\phi)$~\cite{Wilson, Shirasaka, Mauroy1}.

In many cases, the slowest-decaying amplitude associated with $\lambda_1$, which dominates the relaxation dynamics toward the limit cycle, is of interest to us. Retaining only this amplitude, we obtain a pair of phase-amplitude equations in the lowest-order approximation as
$$
{\dot \phi} = \omega + {\bf Z}(\phi) \cdot {\bf P}(t),
\quad
{\dot r} = \lambda r + {\bf I}(\phi) \cdot {\bf P}(t),
\eqno{(5.4)}
$$
where the subscript $1$ is dropped from the amplitude equation.
At the lowest order, the phase is not affected by the amplitude, while the amplitude is modulated by the phase.

By using the amplitude equation in addition to the phase equation, we can analyze the dynamics of the oscillator state around the limit cycle in more detail.
For example, the first-order approximation to the oscillator state is given by ${\bf X}(t) = \chi(\phi(t)) + r(t) {\bf u}_1(\phi(t))$, which would be useful if a more precise oscillator state than the lowest-order approximation $\chi(\phi(t))$ is required in the analysis.
We may also use the amplitude equation for suppressing deviation of the oscillator state from the limit cycle by applying a feedback input to the oscillator in the control problem of synchronization~\cite{Zlotnik,Monga}; for example, if we apply ${\bf P}(t) = - c {\bf u}_1(\phi(t)) r(t)$ with $c > 0$, the amplitude dynamics is given by $\dot{r} = (\lambda - c) r$ and $r$ decays more quickly than the case without the feedback.
Keeping the oscillator state in the close vicinity of the limit cycle by such an additional feedback may allow us to apply stronger input for controlling the oscillator phase to realize desirable synchronized states.

The lowest-order formulation of the phase-amplitude reduction discussed in this subsection corresponds to Method I in Sec.~3. Though the phase is not affected by the amplitudes at this order, it is not the case at the second or higher orders;
for stronger perturbations, more complex dynamics caused by the interaction between the phase and amplitude are expected.
Higher-order generalization of the phase-amplitude reduction and its relation to Method II in Sec.~3 need to be further investigated for a more detailed analysis of the phase-amplitude dynamics.

\subsection{Koopman-operator viewpoint}

In the previous subsection, we mentioned that the recent formulations of the phase-amplitude reduction is closely related to the Koopman-operator approach to nonlinear dynamical systems, which has been intensively studied by Mezi{\'c} and collaborators~\cite{Budic,Mezic} and has attracted much attention. In this subsection, we discuss this relation very briefly.

Suppose a dynamical system $\dot{\bf X} = {\bf F}({\bf X})$ with an exponentially stable limit-cycle solution. We are usually interested in the evolution of the system state ${\bf X}$ in the basin of the limit cycle.
In the Koopman-operator approach, rather than the evolution of the system state ${\bf X}$ itself, we focus on the evolution of the {\it observable} $g$, which is a smooth function that maps the system state ${\bf X}$ to a complex value $g({\bf X})$.
We denote by $\hat\Psi^\tau$ the flow of the system, which maps a system state at time $t$ to a system state at time $t+\tau$ as ${\bf X}(t+\tau) = \hat\Psi^\tau {\bf X}(t)$.
The evolution of an observable $g$ from time $t$ to time $t+\tau$ is described as $g_{t+\tau}({\bf X}) = (\hat U^\tau g_t)({\bf X})$ by using the Koopman operator $\hat U^\tau$ define by
$$ (\hat U^\tau g ) ({\bf X}) = g( \hat \Psi^{\tau} {\bf X} ).
\eqno{(5.5)}
$$
It can easily be shown that $\hat U^0 = I$ (identity), $\hat U^{t+\tau} = \hat U^t \hat U^{\tau}$, and $\hat U^t$ is a linear operator, i.e. $\hat U^\tau(c_a g^a ({\bf X}) + c_b g^b ({\bf X})) = c_a \hat U^\tau g^a ({\bf X}) + c_b \hat U^\tau g^b ({\bf X})$, where $c_{a, b}$ are arbitrary constants and $g^{a, b}$ are two observables. Thus, the observable obeys a linear equation even if the original dynamical system is nonlinear; the cost is that $\hat U^\tau$ is now acting on an infinite-dimensional function space of observables, in contrast to $\hat \Psi^{\tau}$ acting on the finite-dimensional phase space of the oscillator states.

Since $U^\tau$ is a linear operator, it is useful to consider its eigenvalues and eigenfunctions. To see the relation with the phase-amplitude reduction, it is convenient to introduce the infinitesimal generator of $\hat U^\tau$, given by $\hat A g = \lim_{\tau \to +0} ( \hat U^\tau g - g ) / \tau$. By expanding the flow as $\hat \Psi^\tau {\bf X} = {\bf X} + \tau {\bf F}({\bf X}) \tau + O(\tau^2)$ for small $\tau$, we can expand the action of the Koopman operator on $g$ as $(\hat U^\tau g) ({\bf X}) = g( \hat \Psi^\tau ({\bf X}) ) = g( {\bf X} ) + {\rm grad}_{\bf X} g( {\bf X} ) \cdot \tau {\bf F}({\bf X}) + O(\tau^2)$. We thus obtain
$$\hat A = {\bf F}({\bf X}) \cdot {\rm grad}_{\bf X}
\eqno{(5.6)}
$$
and the continuous evolution of the observable $g$ is given by $\dot{g}_t = \hat A g_t$. Note that this generator $\hat A$ of the Koopman operator is exactly the operator that appeared in Eqs.~(3.1) and (5.1) defining the phase and amplitude functions $\phi({\bf X})$ and $r({\bf X})$.

We denote the eigenvalue and the associated eigenfunction of $\hat A$ as $\mu$ and $\varphi_\mu({\bf X})$, respectively, which satisfies $\hat A \varphi_{\mu}({\bf X}) = \mu \varphi_{\mu}({\bf X})$.
There can be infinitely many such eigenvalues and eigenfunctions.
Most importantly, it can be shown that the eigenvalue $i \omega$ and
the Floquet exponents $\lambda_{1, \cdots, n-1}$ of the limit cycle are included in these eigenvalues and they are called the principal eigenvalues~\cite{Mauroy2,Mauroy3}. We denote these eigenvalues as $\mu_0 = i \omega$ and $\mu_l = \lambda_l$ ($l=1, ..., n-1$).

By using the eigenfunctions associated with the principal eigenvalues, we can introduce new coordinates $y_\mu(t) = \varphi_{\mu}({\bf X}(t))$. It then follows that
$$\dot{y}_\mu = \dot{\bf X} \cdot {\rm grad}_{\bf X} \varphi_{\mu}({\bf X}) = {\bf F}({\bf X}) \cdot {\rm grad}_{\bf X} \varphi_{\mu}({\bf X}) = \hat A \varphi_{\mu}({\bf X}) = \mu \varphi_{\mu}({\bf X}) = \mu y_{\mu},
\eqno{(5.7)}
$$
i.e. each $y_\mu$ obeys a simple linear equation $\dot{y}_\mu = \mu y_{\mu}$. Thus, the Koopman eigenfunctions give a global linearization transformation of the system dynamics in the basin of the limit cycle. 
It is also known that analytic observables can be expanded in series by using the principal eigenfunctions~\cite{Mezic}. Therefore, we can reconstruct the observables, including the system state itself, from the new variables $\{ y_\mu(t) \}$.

The amplitude functions $r_{1, ..., n-1}({\bf X})$ defined in the previous subsection are actually principal Koopman eigenfunctions associated with $\mu_{1, ..., n-1}$. Indeed, we have $\hat A r_l({\bf X}) = {\bf F}({\bf X}) \cdot {\rm grad}_{\bf X} r_l({\bf X}) = \mu_l r_l({\bf X})$, which is equivalent to the definition Eq.~(5.1) of $r_l$.
The phase function $\phi({\bf X})$, which satisfies ${\bf F}({\bf X}) \cdot {\rm grad}_{\bf X} \phi({\bf X}) = \omega$ as defined in Eq.~(3.1), is also closely related to the principal Koopman eigenfunction. If we consider an observable $e^{i \phi({\bf X})}$, we have 
$$\hat A e^{i \phi({\bf X})} = {\bf F}({\bf X}) \cdot {\rm grad}_{\bf X} e^{i \phi({\bf X})} = i {\bf F}({\bf X}) \cdot [ {\rm grad}_{\bf X} \phi({\bf X}) ] e^{i \phi({\bf X})} = i \omega e^{i \phi({\bf X})}.
\eqno{(5.8)}
$$
Therefore, $e^{i \phi({\bf X})}$ is also a principal Koopman eigenfunction associated with the eigenvalue $i \omega$.

Thus, we can see a clear correspondence between the Koopman-operator approach, in particular, the global linearization of the system dynamics using the principal Koopman eigenfunctions, and the phase or phase-amplitude reduction of limit-cycle oscillators using the phase and amplitude functions.
Such a relation has recently been clarified by Mauroy, Moehlis, and Mezi{\'c}~\cite{Mauroy3}. They proposed to call the level sets of the amplitude functions as {\it isostables}, which complement the notion of the classical isochrons, i.e. the level sets of the phase function.
Using the idea from the Koopman-operator analysis, they have also proposed a convenient numerical method, called Fourier and Laplace averages, to calculate the phase function $\phi({\bf X})$ and amplitude functions $r_l({\bf X})$~\cite{Mauroy2,Mauroy3}.
The recent developments in the Koopman-operator approach appear to shed new light on the conventional theories of dynamical reduction and further developments in various related fields are expected.

\subsection{Generalization of phase reduction to non-conventional systems}

In Sec. 3, the phase reduction is developed for limit cycles described by ordinary differential equations. Recently, several extensions of the method to non-conventional dynamical systems have been made. We briefly mention some of such attempts here.
\\

(i) {\it Collective oscillations in large populations of globally coupled dynamical elements.}
Large populations of coupled dynamical elements often undergo collective synchronization transition and exhibit stable collective oscillations~\cite{Winfree,Kuramoto,Pikovsky}.
A system of globally coupled noisy phase oscillators is a typical example that exhibits such a collective synchronization transition~\cite{Kuramoto}.
In the large-population limit, the system is represented  by a probability density function $P(\phi, t)$ of the phase, which obeys a nonlinear Fokker-Planck equation. 
This equation has a stable periodic solution satisfying $P(\phi, t) = P(\phi, t+T)$ with period $T$, which represents collective synchronized oscillations of the whole system.
This solution can be considered a limit-cycle solution to the nonlinear Fokker-Planck equation, which should be considered an infinite-dimensional dynamical system.
In Ref.~\cite{Kawamura1}, an equation of the form~(3.3) for the collective phase of the nonlinear Fokker-Planck equation is obtained and synchronization between a pair of such collectively oscillating systems is analyzed.
In Ref.~\cite{Kawamura2}, a similar phase equation is also derived for the collective phase of a system of globally coupled noiseless phase oscillators with sinusoidal coupling and Lorentzian frequency heterogeneity, which also exhibits collective synchronization transition, in the large-population limit by using the Ott-Antonsen ansatz~\cite{Ott}.
Such dynamical reduction for collective oscillations, which we called collective phase reduction, would be useful in analyzing collective oscillations in large populations of dynamical elements at the macroscopic level.
For example, macroscopic rhythms arising in a mathematical model of neural populations have been analyzed by using a similar approach~\cite{Kotani0}.

(ii) {\it Collective oscillations in networks of coupled dynamical systems.} Stable collective oscillations in networks of coupled dynamical systems often play important functional roles in the real-world systems such as neuronal networks and power grids.
When the collective oscillations correspond to a limit-cycle orbit in the high-dimensional phase space of the network, we can develop a phase reduction theory, namely, we can define a phase of the collective oscillation of the network and represent the network dynamics by a single phase equation.
In Ref.~\cite{Kori}, such a phase reduction theory has been developed for collective oscillations in a network of coupled phase oscillators, where the network topology can be arbitrary. The sensitivity functions of the individual elements, characterizing how tiny perturbations applied to each element affect the collective phase of the network, have been derived.
In Ref.~\cite{Nakao}, the theory has further been generalized to arbitrary networks of coupled heterogeneous dynamical elements, where the dynamics of individual elements can also be arbitrary.
As long as the network exhibits stable collective limit-cycle oscillations as a whole, the network dynamics can be reduced to a single phase equation for the collective phase and, for example, synchronization between a pair of such networks can be analyzed by using the same classical methods as those for ordinary low-dimensional oscillators.
The interplay between the individual dynamics of the elements and network topology can lead to strongly heterogeneous sensitivities of the individual dynamical elements; for example, some special elements in the network can possess much stronger sensitivity than the other elements.
Such information would be useful, for example, in developing a method to control the collective oscillations of the network by external input signals.

(iii) {\it Delay-induced oscillations.} Delay differential equations are used in various fields of science and engineering, e.g. in mathematical models of biological oscillations caused by delayed feedback.  Since delay-differential equations are infinite-dimensional dynamical systems that depend on their past states, the conventional method of phase reduction for ordinary differential equations cannot be used for them. In Refs.~\cite{Pyragas,Kotani}, phase reduction methods for limit-cycle oscillations in delay-differential equations with a single delay time have been developed. The difficulty is that the oscillator state is given by a function representing its past history and not simply by an ordinary finite-dimensional vector. In Ref.~\cite{Kotani} by Kotani {\it et al.},
the adjoint equation for the sensitivity function is derived on the basis of the abstract mathematical theory of functional differential equations, while in Ref.~\cite{Pyragas} by Novi{\v c}enko and Pyragas, a physically more intuitive approach has been taken to derive the phase equation. 
Both of these results are consistent with each other and enable us to reduce infinite-dimensional delay-differential equations to a one-dimensional phase equation, thereby providing a powerful method to analyze synchronization properties of delay-induced oscillations under weak perturbations. 
Though Refs.~\cite{Pyragas,Kotani} treated oscillator models with a single delay time, one should be able to generalize the theory to allow multiple or distributed delay times. 

(iv) {\it Oscillations in spatially extended systems.} Nonlinear oscillatory dynamics in real-world systems often arise in spatially extended systems. The target or spiral waves in chemical or biological systems described by reaction-diffusion equations~\cite{Mikhailov} and oscillatory fluid flows described by Navier-Stokes equations~\cite{Taira} are typical examples. Similar to the delay-differential equations, these partial differential equations are also infinite-dimensional dynamical systems and the conventional method of phase reduction cannot be applied to them; even so, as long as the system exhibits stable limit-cycle oscillations, one should be able to assign a phase value to the system state described by a spatially extended field variable, and derive a phase equation describing the whole system.
In Ref.~\cite{Kawamura3}, a phase reduction method for a fluid system has been developed, where a phase equation describing oscillatory convection in a Hele-Shaw cell is derived from the fluid equation, and synchronization between two coupled convection cells are analyzed. The sensitivity function for the phase is represented by a two-dimensional field with characteristic localized structures, representing strong spatial heterogeneity of the response properties of the oscillatory convection to applied perturbations.
In Ref.~\cite{Nakao2}, a general phase reduction method for reaction-diffusion systems exhibiting rhythmic spatiotemporal dynamics has been developed and typical rhythmic patterns, such as oscillating spots, target waves, and spiral waves arising in the FitzHugh-Nagumo reaction-diffusion system, are analyzed using the reduced phase equation. The notion of phase functional, which generalizes the phase function in Sec. 3, is introduced, and the sensitivity function of the phase is defined as a functional derivative of the phase functional. An adjoint partial differential equation describing the sensitivity function of the system is also derived. The sensitivity function exhibits strong spatial heterogeneity also in this case, reflecting spatiotemporal structures of the rhythmic patterns.
It also seems possible to develop similar phase-reduction theories for other types of partial differential equations exhibiting spatiotemporally rhythmic patterns, and such theories would be useful in the analysis and control of rhythmic spatiotemporal patterns.

(v) {\it Hybrid nonlinear oscillators.} Hybrid, or piecewise-smooth dynamical systems, consisting of smooth dynamics described by ordinary differential equations and discontinuous jumps connecting them, are often used in mathematical models of real-world phenomena; the simplest example is the bouncing ball with discontinuity in the vertical velocity. Stable limit cycles can also arise in such systems, for example, in mathematical models of passively walking robots or electric oscillators with switching elements. The conventional phase reduction method cannot be applied to such hybrid systems because of the discontinuity. However, again, as long as the system exhibits stable limit-cycle oscillations, one should be able to define a phase and derive a phase equation. In Refs.~\cite{Shirasaka2,Park1}, phase equations for hybrid limit-cycle oscillators have been derived. A generalized adjoint equation with discontinuous jumps is derived and the sensitivity function for the phase is obtained. Due to the jumps, the sensitivity function can possess non-smooth discontinuities, which can lead to nontrivial synchronization dynamics that cannot occur in smooth dynamical systems.

(vi) {\it Strongly perturbed and non-autonomous oscillators.} In this  article and in all generalizations mentioned above, the property of the oscillatory system, whether it is low-dimensional or infinite-dimensional, is assumed to be constant, namely, the limit-cycle orbit, natural frequency, and phase sensitivity function do not vary with time.
However, this may not be the case in some realistic problems, for example, when we consider biological oscillators that are subjected to slow but strong effect of parametric modulation due to circadian rhythms in addition to small perturbations coming from other oscillators.
In such cases, it is convenient to regard the oscillator as a non-autonomous system whose parameter varies slowly with time while being subjected to fast small perturbations.
In Refs.~\cite{Kurebayashi1,Kurebayashi2,Park2}, the phase reduction method has been generalized to such non-autonomous oscillators.
By considering a continuous family of limit cycles parameterized by the slowly varying input, a closed phase equation for the oscillator can be derived, which is characterized by the instantaneous frequency and two sensitivity functions.
One of the sensitivity functions is the ordinary phase sensitivity function to small perturbations, and the other is a newly introduced sensitivity function to characterize the effect of deformation of the limit cycle due to slow parametric variation.
Using the extended phase equation, nontrivial synchronization of a limit-cycle oscillator subjected to strong periodic forcing~\cite{Kurebayashi1} and between a pair of coupled oscillators subjected to parametric variation~\cite{Park2} have been analyzed.
As emphasized in~\cite{Suprunenko}, non-autonomous systems can be essentially important in modeling real biological oscillations and further studies into this direction, e.g., collective phase synchronization of non-autonomous oscillators, would provide useful information for understanding real biological systems.

In this subsection, we have briefly described some of the recent extensions of the phase reduction to non-conventional rhythmic systems.
The simplicity and flexibility of the phase reduction, whose key ideas have been discussed in the main text, allow us to extend the method to a wide class of nonlinear oscillatory dynamical systems.
Moreover, once individual systems are reduced to phase equations, synchronization between coupled oscillatory systems of different types, e.g., networked systems and spatially extended systems, can readily be analyzed within the framework of the phase reduction. Such an approach would be useful in analyzing complex multiscale systems consisting of oscillatory subsystems of different types like neuronal systems. 
Further extensions of the phase-reduction approach would also be possible for other classes of nonlinear oscillatory systems, including stochastic, chaotic, and even quantum systems~\cite{Kato}.

Finally, though not considered in the present paper, data-driven inverse-type approaches, which infer reduced dynamical equations from experimental data without knowing the precise mathematical models of the oscillatory phenomena, are becoming increasingly important recently. As discussed in Sec. 4, reduced equations can often be obtained without the detailed knowledge of individual systems in the dynamical reduction theory.
This implies that one can infer appropriate reduced equations only from the experimental data of complex oscillatory phenomena and use them to predict their dynamics without knowing their precise physical, chemical, or biological origins.
See, for example, Refs.~\cite{Tokuda,Kralemann,Stevanovska,Aoyagi} for such inverse approaches to infer reduced phase equations from complex oscillatory phenomena.\\

{\bf Acknowledgments:} We would like to thank Prof. Aneta Stevanovska, Prof. Tomislav Stankovski, and Prof. Peter McClintock for kindly inviting us to write this overview. We also thank Dr. Yoji Kawamura for fruitful collaboration and discussions on the dynamical reduction of nonlinear oscillatory systems.\\

{\bf Author contributions:} Y. K. formulated the theories of center-manifold and phase reduction for limit-cycle oscillators in the main text. H. N. discussed the phase-amplitude reduction and relation to the Koopman-operator approach in the appendix. Both authors read and approved the manuscript.\\

{\bf Funding:} H.N. is supported by the JSPS KAKENHI Grant Numbers JP16K13847, JP17H03279, 18K03471, and JP18H03287.\\

\end{document}